\documentclass[acmsmall]{acmart}
\usepackage{algorithm}
\usepackage{subcaption}
\usepackage{algpseudocode}
\usepackage{mathrsfs}
\usepackage{bm}

\AtBeginDocument{%
  \providecommand\BibTeX{{%
    \normalfont B\kern-0.5em{\scshape i\kern-0.25em b}\kern-0.8em\TeX}}}

\setcopyright{acmlicensed}
\copyrightyear{2024}
\acmYear{2024}
\acmConference[Preprint, Under Review]{Under Review}{June}{2024}

\renewcommand\footnotetextcopyrightpermission[1]{}
\begin{document}
\citestyle{acmauthoryear}

\title{DirectL: Efficient Radiance Fields Rendering for 3D Light Field Displays}

\author{Zongyuan Yang}
\authornote{The authors contributed equally to this work.}
\author{Baolin Liu}
\authornotemark[1]
\author{Yingde Song}
\authornotemark[1]
\author{Yongping Xiong}
\authornote{First corresponding author}
\affiliation{%
    \\
  \institution{State Key Laboratory of Networking and Switching Technology, \\Beijing University of Posts and Telecommunications}
  \country{China}
}

\author{Lan Yi}
\affiliation{%
   \institution{Beijing University of Posts and Telecommunications}
  \country{China}}

\author{Zhaohe Zhang}
\author{Xunbo Yu}
\authornote{Second corresponding author}
\affiliation{%
\\
 \institution{State Key Laboratory of Information Photonics and Optical Communications, \\Beijing University of Posts and Telecommunications}
  \country{China}}

\renewcommand{\shortauthors}{Yang, Liu and Song, et al.}

\begin{abstract}
  Autostereoscopic display technology, despite decades of development, has not achieved extensive application, primarily due to the daunting challenge of three-dimensional (3D) content creation for non-specialists. The emergence of Radiance Field as an innovative 3D representation has markedly revolutionized the domains of 3D reconstruction and generation. This technology greatly simplifies 3D content creation for common users, broadening the applicability of Light Field Displays (LFDs). However, the combination of these two technologies remains largely unexplored. The standard paradigm to create optimal content for parallax-based light field displays demands rendering at least 45 slightly shifted views preferably at high resolution per frame, a substantial hurdle for real-time rendering. We introduce DirectL, a novel rendering paradigm for Radiance Fields on autostereoscopic displays with lenticular lens. We thoroughly analyze the interweaved mapping of spatial rays to screen subpixels, precisely determine the light rays entering the human eye, and propose subpixel repurposing to significantly reduce the pixel count required for rendering. Tailored for the two predominant radiance fields—Neural Radiance Fields (NeRFs) and 3D Gaussian Splatting (3DGS), we propose corresponding optimized rendering pipelines that directly render the light field images instead of multi-view images, achieving state-of-the-art rendering speeds on autostereoscopic displays. Extensive experiments across various autostereoscopic displays and user visual perception assessments demonstrate that DirectL accelerates rendering by up to 40 times compared to the standard paradigm without sacrificing visual quality. Its rendering process-only modification allows seamless integration into subsequent radiance field tasks. Finally, we integrate DirectL into diverse applications, showcasing the stunning visual experiences and the synergy between Light Field Displays and Radiance Fields, which unveils tremendous potential for commercialization and practical applications. \href{https://direct-l.github.io/}{\textbf{DirectL Project Homepage: https://direct-l.github.io/}}
\end{abstract}

\ccsdesc[500]{Computing methodologies~Rendering}
\ccsdesc[500]{Computing methodologies~3D imaging}
\ccsdesc[500]{Hardware~Displays and imagers}

\keywords{Radiance fields, Light field displays}

\begin{teaserfigure}
  \begin{subfigure}[h]{0.48\textwidth}
    \includegraphics[width=\textwidth, height=0.9\textwidth]{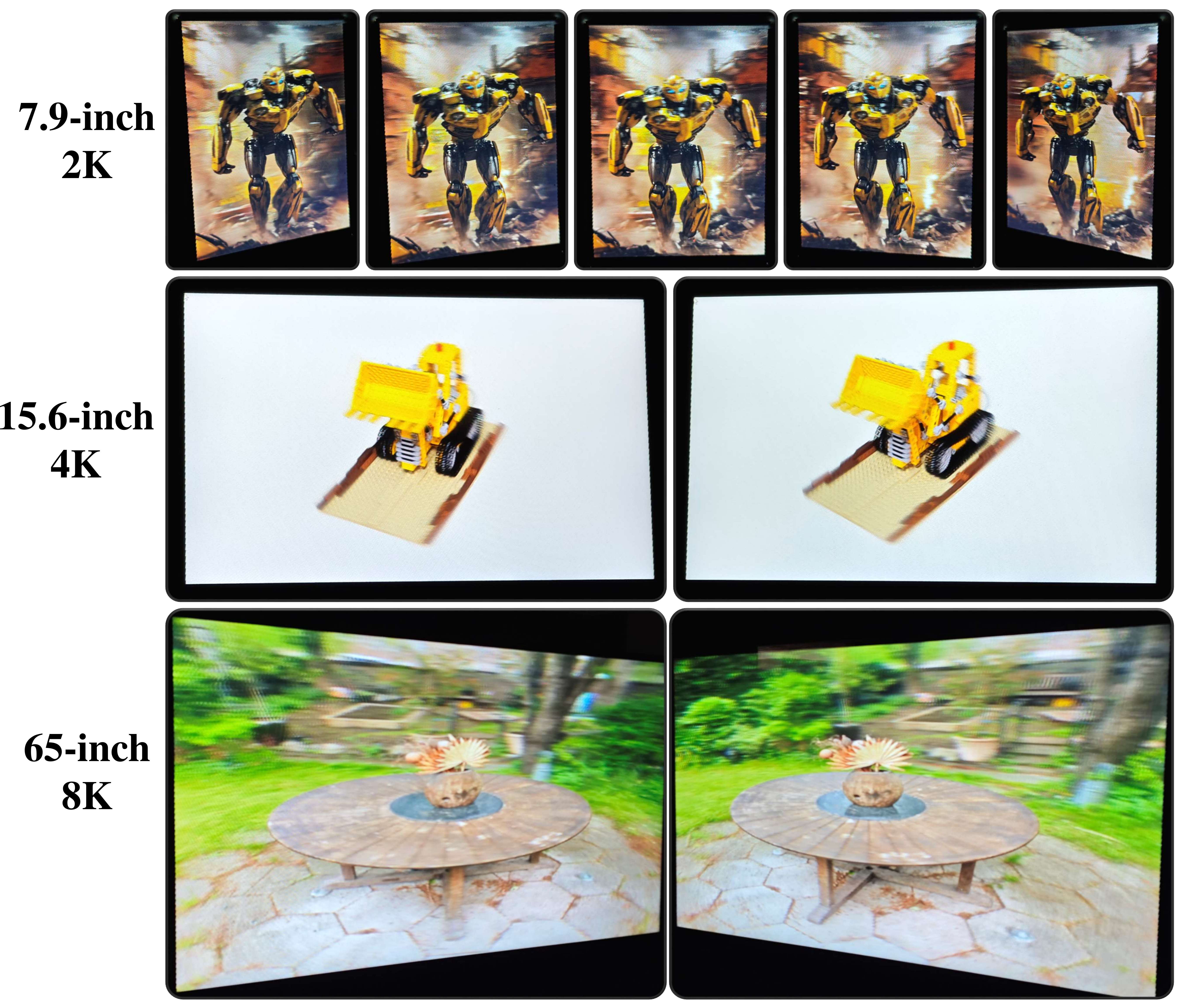}
    \caption{The synergistic combination of radiation fields and light field displays, leveraging our proposed DirectL, facilitates the realization of visually compelling, user-friendly, rich and photorealistic naked-eye 3D experiences.}
    \label{fig:teaser1}
  \end{subfigure}
  \hfill
  \begin{subfigure}[h]{0.5\textwidth}
    \includegraphics[width=\textwidth, height=0.9\textwidth]{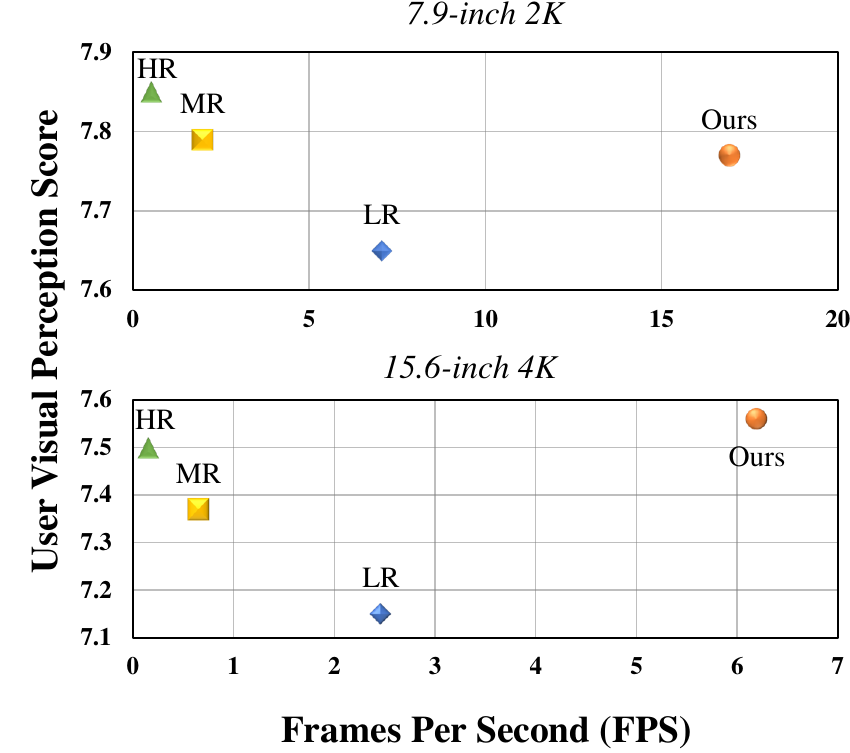}
    \caption{Comparison of DirectL and the current standard paradigm in terms of FPS and user visual perception quality on two mainstream consumer-level light field displays.}
    \label{fig:teaser2}
  \end{subfigure}
  \caption{We first combine radiance fields and light field displays at a fundamental level, enabling efficient rendering of radiance field methods on light field screens. This mitigates the long-standing challenge hindering the widespread application of light field displays: the paucity of 3D display content. (a) showcases the compelling naked-eye 3D effects achieved through this combination. The first row illustrates the 2D image of \textit{Bumblebee} lifted into 3D and rendered effectively through the DirectL framework. The second and third rows display the classic \textit{Lego} from \cite{mildenhall2021nerf} and \textit{Garden} from \cite{barron2022mip} but in naked-eye 3D. (b) illustrates that DirectL significantly enhances rendering speed while maintaining visual quality. Due to the absence of quantitative perception metrics for evaluating 3D display effects, we conduct a user study with 30 unbiased participants (see Appendix \ref{User Study}). \href{https://direct-l.github.io/}{\textbf{DirectL Project Homepage}} includes more demos and VR videos to replicate binocular parallax for those without access to a light field screen.}
  \label{fig:teaser}
\end{teaserfigure}

\maketitle

\section{Introduction}
Light Field Displays (LFDs) can reproduce the light rays' distribution of 3D scenes, providing viewers with a realistic stereoscopic vision that includes binocular disparity and occlusion effects \cite{urey2011state, wu2017light, shen2023virtual}. There are many variants of LFDs, among which Autostereoscopic Displays (ADs) have fast-paced development and widespread attention in recent years due to their capabilities to operate without additional wearables and support multi-user viewing simultaneously\cite{matusik20043d, dodgson2005autostereoscopic, perlin2000autostereoscopic, efrat2016cinema}. ADs primarily divide into parallax barrier displays and lenticular lens displays. A notable product applying parallax barrier display is the Nintendo 3DS \cite{Stuart2011}. However, due to low visual quality and limitations on viewing angles, the majority of stereoscopic devices on the market are now based on lenticular lens displays \cite{shen2023virtual}, which are the focus of this paper. Lenticular lens displays involve affixing a slightly tilted array of lenticular sheets in front of an LCD panel, allowing light rays to be refracted to different spatial positions, thereby creating horizontal parallax for the human eye and producing a stereoscopic visual effect. With increasing LCD panel resolution and decreasing costs, common consumers can now purchase products based on lenticular lens display technology, such as Looking Glass \cite{LookingGlass2024a} and X-real \cite{ZXReal2024}. It is worth mentioning that Apple Vision Pro's EyeSight \cite{kim2023optical} also utilizes the lenticular lens displays to project the correct viewing angle of the virtual eye images onto each person observing it.

Despite the increasing accessibility of light field displays to common users, seamlessly projecting nearby scenes onto these screens with interactions remains a formidable challenge. Two main challenges lie ahead: 1) The difficulty for non-specialists to create 3D display content. 2) The immense computational cost of rendering. For the second point, as spatial viewpoints share pixels on a single LCD panel, improving the perceived resolution for each viewpoint requires maximizing the LCD panel's resolution, which means more pixels need to be rendered. For optimal display quality, a standard rendering paradigm \cite{shen2023virtual, fink2023efficient, van1999image} for a 15.6-inch 3D display is to first render 60 slightly offset views at a resolution of 2160 $\times$ 3840, then interweave them into an encoded image. Considering computational complexity, an acceptable configuration is 60 views at 450px $\times$ 800px per view. However, this still requires 2.6 times the original resolution, and it leads to significant visual quality degradation as direct interpolation causes ghosting artifacts\cite{woods2012crosstalk}  (see Sec. \ref{Baseline Comparisons}). Consequently, a natural question arises: Can we circumvent the rendering of extraneous pixels without compromising visual fidelity? While rendering from existing 3D assets like meshes is a feasible approach, creating meshes from 2D resources poses significant difficulty for non-specialists. As an innovative 3D representation, Radiance Fields \cite{mildenhall2021nerf, kerbl20233d} enable high-quality scene reconstruction and arbitrary viewpoint rendering from merely captured video, with the latest large reconstruction models \cite{hong2023lrm, tochilkin2024triposr, tang2024lgm} capable of reconstructing entire objects from a single frontal view. To adapt to the traditional graphics rendering pipeline for downstream tasks \cite{tang2023delicate, guedon2023sugar}, Radiance Fields often need to be exported into meshes through marching cubes algorithms \cite{shen2021deep, lorensen1998marching} . However, the exported meshes still face issues such as geometric errors and topological chaos, which greatly limits their application scope. In contrast, Radiance Fields only require rendering the necessary viewpoints for LFDs. This makes the application of Radiance Fields on LFDs more convenient and efficient. Tensor4D \cite{shao2023tensor4d} showcases offline rendering of 4D Radiance Fields on light field displays for reconstruction effects. In \cite{stengel2023ai}, monocular RGB video is captured and lifted into triplane-based NeRFs using NVIDIA RTX 6000 GPUs, then locally rendered into multi-view images and interweaved for light field displays. Despite extensive engineering optimizations and leveraging high-performance RTX 4090 GPU, the method following the standard paradigm still exhibits low rendering performance. 

Currently, existing works \cite{stengel2023ai, shao2023tensor4d, tu2024tele} applying Radiance Fields to LFDs merely connect the two through high-level SDK-like integration, without thoroughly exploring their principled combination. Radiance Fields on LFDs still face the aforementioned low rendering efficiency issue. Therefore, our \textbf{motivation} is to address the challenges in content creation and rendering efficiency for light field displays. We particularly focus on how to combine Radiance Fields with LFDs at a lower level for efficient Radiance Fields rendering on LFDs. Through this combination, we expect to reduce the long-standing complexity of creating content for LFDs while significantly improving radiance field rendering efficiency on LFDs, thereby providing users with a more realistic and immersive stereoscopic visual experience. We also hope to break down the barrier between the two fields, attracting more researchers to join and promote the popularization and development of light field displays. 

In this paper, we propose a novel Radiance Fields rendering paradigm called DirectL, tailored for Light Field Displays, that effectively addresses the two aforementioned issues. First, we conduct an in-depth analysis of the interleaving process, i.e., the mapping between pixels in the multi-view images and pixels in the encoded image, to determine which pixels will be displayed and which will be wasted. This is equivalent to identifying the light rays in space that need to be rendered. In the standard rendering paradigm, over 90\% of the pixels in the multi-view images are wasted. However, even if the standard paradigm could somehow render only the required pixels, it would still necessitate rendering three times the original resolution for light field displays (see Sec. \ref{Lenticular Lens Based 3D Display System}). To address this, we propose repurposing the idle subpixels at nearby positions in the same viewpoint image for the corresponding pixel position in the encoded image. This strategy further reduces the required rendered pixel count by approximately 55\% while negligibly affecting visual quality. Next, we propose optimized rendering pipelines for the two predominant radiance field representations, Neural Radiance Fields (NeRFs) \cite{muller2022instant, hedman2021baking} and 3D Gaussian Splatting (3DGS) \cite{kerbl20233d}, to achieve state-of-the-art rendering speeds on LFDs. Specifically, for NeRFs volume rendering, we compute the required light ray directions and arrange the inference order such that the red channel aligns with the encoded image. Based on pre-computed indices, we perform efficient GPU-accelerated reordering of the green and blue channels, reshaping the output into an encoded image frame that can be directly displayed on the light field displays. For 3D Gaussian Splatting (3DGS) tile-based splatting rasterization, inspired by \cite{mueller1996fast} and \cite{hubner2006multi}, we transform the rendering process from object-order to image-order, or more specifically, to ray-order. This conversion to ray-order rendering enables the execution of a rendering workflow consistent with NeRFs for generating light field images.  Finally, we integrate the DirectL into multiple downstream radiance field-based reconstruction and generation tasks \cite{xu2023gaussian, wu20234d, xu20234k4d, tochilkin2024triposr}, demonstrating that radiance fields can greatly enrich 3D display content, simplify production, and enable more real-time interactions, revealing the rationale for combining light field displays and radiance fields, and the potential opportunities and revolutions this may bring. 

In summary, we make the following contributions:
\begin{itemize}
    \item The introduction of DirectL, a novel Radiance Fields rendering paradigm tailored for Light Field Displays, which includes specialized rendering pipelines for NeRFs and 3DGS, enabling state-of-the-art rendering performance on LFDs.
    \item An optimization interleaving method that repurposes idle subpixels at nearby positions within the same view, further reducing the required rendering pixels by approximately 55\% with negligible visual quality degradation, particularly beneficial for image-order radiance fields rendering.
    \item The integration of DirectL with downstream radiance field tasks, showcasing radiance fields' potential to enrich 3D display content, simplify content creation, and enable real-time interactions, while unveiling the synergies and transformative opportunities arising from combining light field displays and radiance fields.
\end{itemize}

We evaluate our proposed DirectL on lenticular autostereoscopic displays of 7.9 inches with 2K resolution, 15.6 inches with 4K resolution, and 65 inches with 8K resolution. Comprehensive experimental results demonstrate that, in comparison to the standard paradigm, DirectL can achieve up to 40$\times$ rendering speedup while maintaining equivalent visual quality. This makes DirectL \textbf{the first method} capable of real-time high-quality radiance fields rendering ($\geq$ 25 FPS) on the 48-viewpoint 2K resolution light field display.

\section{RELATED WORK}
In this section, we first briefly summarize traditional approaches for light field image creation, then discuss work on efficient rendering for lenticular-based light field displays. Next, we review recent developments in radiance field research. Finally, we discuss the synergistic relationship and necessity of combining radiance fields and light field displays.
\subsection{Efficient Rendering for 3D Light Field Displays}
The content displayed on light field displays can also be referred to as light field image. The creation of light field images has been a longstanding challenge \cite{gortler2023lumigraph, halle1998multiple, levoy2006light, kalantari2016learning, shi2017near, zhou2021review, shen2023virtual, levoy2023light, gavane2024light}. The most common method for capturing light field images in the real world involves recording light field information from multiple viewpoints simultaneously using an array of cameras \cite{li20153d}. Plenoptic cameras produce light field images by capturing spatial and angular information of light in a scene through an array of microlenses \cite{ng2005light}. However, the high cost of hardware and the scarcity of content have been the primary obstacles to the widespread adoption of light field displays \cite{algorri2016liquid, urey2011state}.

Most light field displays are designed to display a frontal light field \cite{shen2023virtual}. Consequently, light field images can also be created by rendering a frontal light field from existing 3D assets, such as meshes. Nevertheless, the computational cost of rendering a light field image is substantial. Given a fixed resolution of the LCD panel, there is a trade-off between spatial and angular resolutions. With advancements in LCD panel resolution and manufacturing techniques, it is now available to have a 65-inch light field display \cite{LookingGlass2024c, ZXReal2024} with an LCD panel resolution of 7680 $\times$ 4320, capable of displaying 96 viewpoints across an 80-degree field of view. To render a light field image for such a display, it is necessary to render images of at least 1.5K resolution or higher for each viewpoint to ensure an acceptable display quality. The required resolution for a single viewpoint is related to the complexity of the 3D scene and the optical structure of the display. Within the displayable depth \cite{he2024assessment} and equivalent single-viewpoint resolution supported by the optical structure, the more complex the 3D scene, the higher the single-viewpoint resolution required for better display effects \cite{he2024assessment, nam2017flat, li20153d}. In theory, rendering images for each single viewpoint at the same resolution as the LCD panel and interlacing them would provide the best display quality \cite{he2024assessment, nam2017flat}. However, in practice, due to the significant computational cost, compromises must be made on the resolution of the single viewpoints. 

To address this, various methods have been developed to improve the traditional rendering pipeline for efficient rendering on light field screens.
\citet{hubner2006multi} introduces a multi-view point-rendering algorithm, which performs per-fragment blending and splat projection calculations in a single pipeline pass, thereby streamlining the multi-view rendering process for point-based visualizations in autostereoscopic displays. \citet{nozick2010multi} presents a viewpoint-by-view rendering method that uses geometric shaders to copy primitives across views. This reduces rendering time but greatly increases VRAM usage, limiting it to small scenes. \citet{marrs2017real} introduces an efficient view-independent rasterization technique for multi-view rendering that by transforming polygonal models into point clouds specialized for multiple views. \citet{guan2020parallel} present a parallel multi-view polygon rasterization algorithm for lenticular-based light field displays that exploits the coherence of rasterization calculations across views, enabled by a GPU-driven hierarchical soft rendering pipeline. Recent work \cite{fink2023efficient} has made progress  in minimizing the rendering computation for individual viewpoints. \citet{fink2023efficient} introduce custom projective mappings tailored for lenticular-based light field displays to minimize the rendering of ineffective pixels. However, this approach has several limitations. Firstly, the method of reducing the resolution for single viewpoints requires extensive upsampling interpolation during the interlacing process, leading to blurring artifacts. Moreover, the method \cite{fink2023efficient} is only tested on a 7.9-inch 2K light field screen displaying simple single objects. Its strategy of rendering "flattened" multi-view images to decrease rendering computation sacrifices display depth in complex 3D scenes, thereby severely impacting the 3D display effect \cite{nam2017flat}.

\begin{figure*}[!h]
  \centering
  \begin{subfigure}[h]{\textwidth}
    \includegraphics[width=\textwidth]{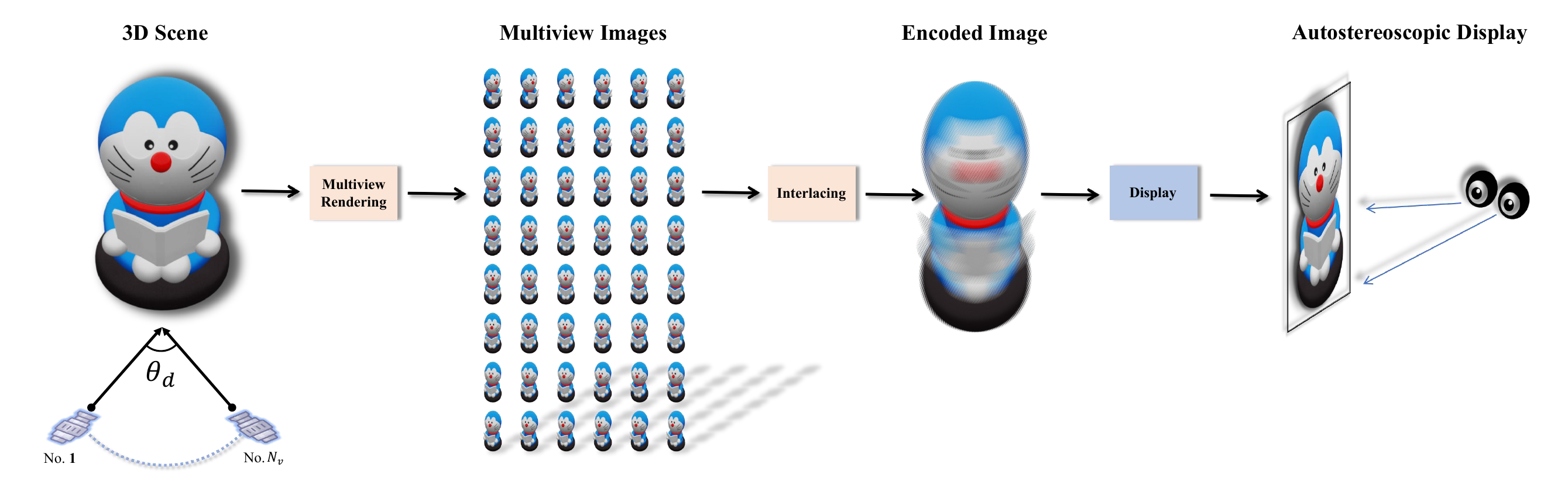}
    \caption{Standard paradigm to render light field images for 3d displays }
    \label{fig:display_sub1}
  \end{subfigure}

  \begin{subfigure}[h]{0.47\textwidth}
    \includegraphics[width=\textwidth]{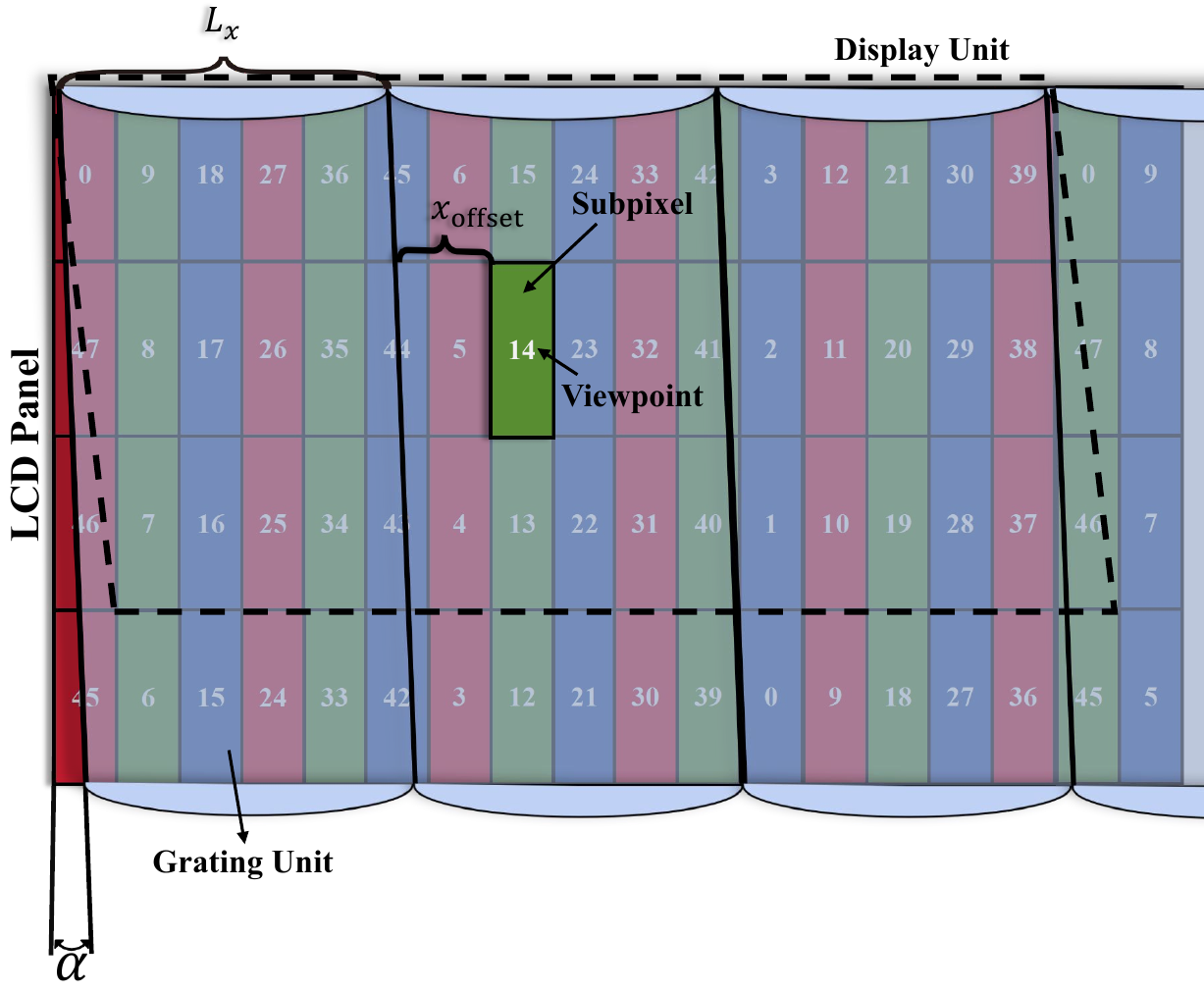}
    \caption{$\alpha = \arctan(-\frac{1}{27})$, $L_{\rm x} = 5\frac{1}{3}$, $K_{\rm offset} = 0 $, $N_{\rm v} = 48$, viewpoint number matrix $M_{\rm v}$ diagram}
    \label{fig:display_sub2}
  \end{subfigure}
  \hfill
  \begin{subfigure}[h]{0.47\textwidth}
    \includegraphics[width=\textwidth]{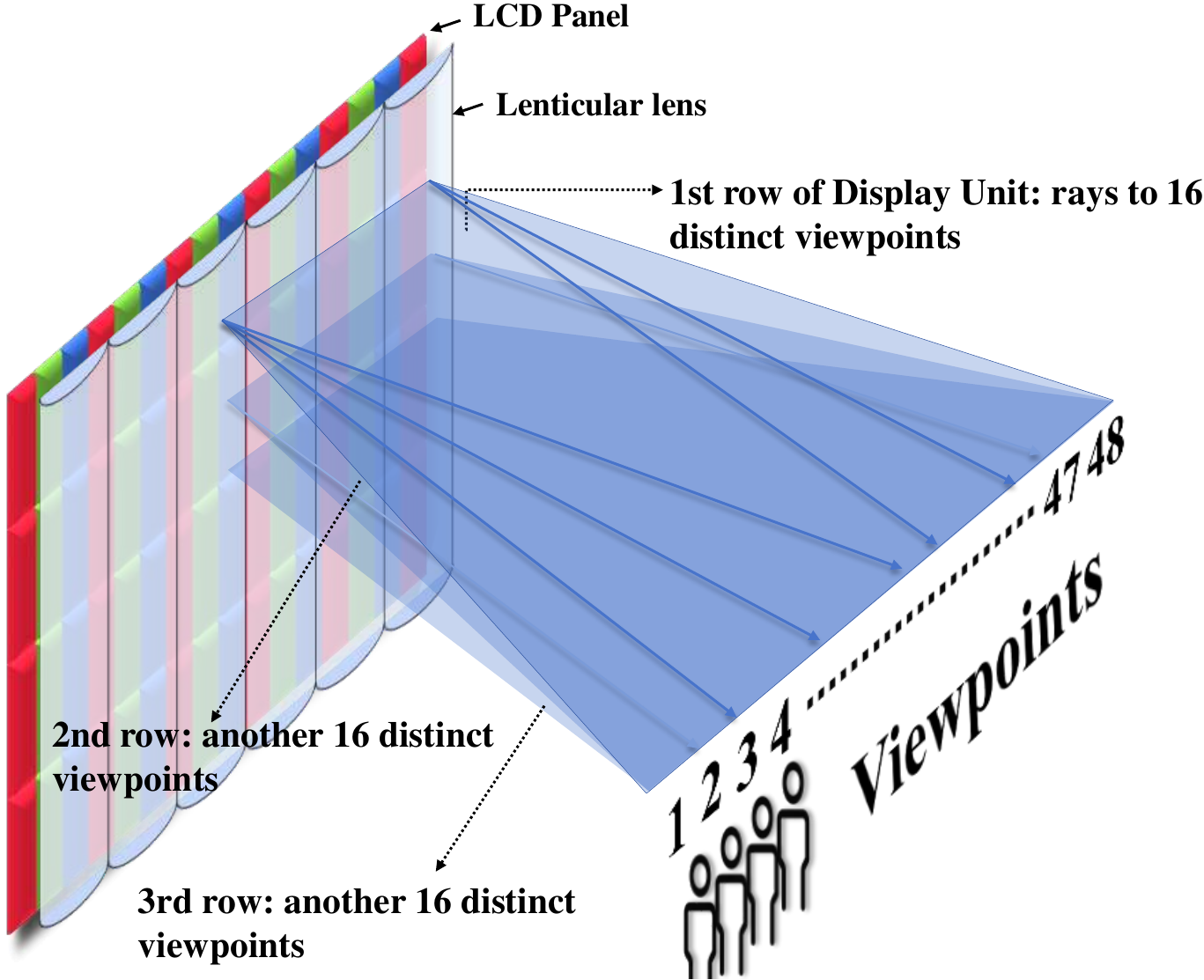}
    \caption{Spatial arrangement of viewpoints at the best viewing plane}
    \label{fig:display_sub3}
  \end{subfigure}

  \caption{The overall of standard lenticular lens based 3D display system.}
  \label{fig:display}
\end{figure*}

\subsection{Radiance Fields Rendering}
Theoretically, any method capable of rendering frontal light fields is applicable to light field displays, thus radiance field-based methods are no exception. Radiance field is a burgeoning research area with a wealth of significant contributions in recent years.  Our focus is on foundational works pertinent to light field display applications.  Neural Radiance Field (NeRF) introduced by \cite{mildenhall2021nerf} employs Multi-Layer Perceptrons (MLP) to learn volumetric representations of a 3D scene and utilizes volumetric ray-marching for rendering, which is computationally intensive due to the multiple MLP queries required per pixel.  Many subsequent works, such as \cite{muller2022instant, chen2022tensorf, fridovich2022plenoxels, chen2023mobilenerf, reiser2023merf, duckworth2023smerf, hedman2021baking, duckworth2023smerf, yariv2023bakedsdf}, have aimed to enhance rendering efficiency and maintain image quality.   \citet{hedman2021baking} proposes Sparse Neural Radiance Grid (SNeRG), which bakes trained NeRFs in a sparse voxel grid, enabling real-time rendering across platforms. Following iterations have further refined this concept with strategies like textured polygons \cite{chen2023mobilenerf}, hybrid volumetric parameterization \cite{reiser2023merf}, and hierarchical model partitioning \cite{duckworth2023smerf}. Instant-ngp \cite{muller2022instant} notably accelerates computation and enhances image quality by incorporating multi-resolution hash encoding, becoming one of the most notable methodologies within NeRFs. The recent advent of 3D Gaussian Splatting (3DGS) \cite{kerbl20233d} ushers a new era for radiance fields, representing scenes through explicit 3D Gaussians and achieving state-of-the-art visual quality and real-time rendering via tile-based differentiable splatting rasterization. The proliferation of studies based on 3DGS improvements (\cite{fan2023lightgaussian, niedermayr2023compressed, huang20242d}) or applications (\cite{xu20234k4d, wu20234d, yang2023deformable, xu2023gaussian, xu2024grm, tang2024lgm}) have emerged ceaselessly \cite{fei20243d, chen2024survey}, showcasing the field's dynamic progress and potential. 

Reconstructing 3D scenes with radiance fields naturally reconstructs the light field, making it suitable for light field displays. Furthermore, radiance field methods have greatly simplified the the creation of light field images for non-specialists—combining generation methods \cite{hong2023lrm, tochilkin2024triposr, zhang2024gs} enables quick creation of a wealth of light field images. Despite the potential, research on light field displays and radiance fields have run parallel without intersection. While existing works have demonstrated the possibility of combining these fields at a high level, a deeper integration is desired. This paper aims to organically intertwine light field display principles with radiance field methods, intersecting two parallel fields.

\section{PRELIMINARIES}
In this section, we introduce the preliminaries necessary before presenting DirectL. Firstly, we explain the lenticular lens-based 3D display system from a multi-view rendering standpoint, constituting the theoretical groundwork for DirectL's effective display on autostereoscopic 3D screens. Subsequently, we summarize the Neural Radiance Fields (NeRFs) and 3D Gaussian Splatting (3DGS) approaches, pivotal for DirectL's proficient reconstruction of 3D scenes' light fields. 

\subsection{Lenticular Lens Based 3D Display System}
\label{Lenticular Lens Based 3D Display System}
The human eye perceives depth through binocular disparity, and the rendering of multi-view images is essentially to generate a sequence of disparity images. The process of selecting specific subpixels from this sequence and arranging them into a structured pattern results in an encoded image—a subtype of light field images—referred to as interleaving.  Displaying the encoded image on the LCD panel of an autostereoscopic display, the controlling effect of the lenticular lens grating causes the subpixels' emitted light to form distinct viewpoint regions in space. An observer, with each eye positioned in these separate regions, experiences a stereoscopic image. Thus, the lenticular lens-based 3D display system encompasses three stages: multi-view rendering, interleaving, and optical display, as illustrated in Fig. \ref{fig:display}. 

\textbf{Multiview Rendering}. As depicted in Fig. \ref{fig:display_sub1}, $N_{\rm v}$ cameras are evenly spaced along an arc \footnote{
Off-axis perspective with straight lines is another option.}, each targeting the central object with their focal planes converging at the circle's center. The angular span $\theta_{\rm d}$ between the outermost cameras defines the display's field of view, determined by the optical structure of the lens. Taking, for instance, our experimental 65-inch 8K resolution 3D display, it features $N_{\rm v} = 96$ viewpoints with a $\theta_{\rm d}$ of 80°. In practice, with a fixed $\theta_{\rm d}$, rendering additional viewpoints does not impair the display; instead, it smoothens the transition between views \cite{nam2017flat, shen2023virtual}. This is because an increased number of disparity images provides a surplus of light field information, and the interleaving process accurately maps the subpixels from any length of disparity image sequence onto the encoded image. 

\textbf{Interlacing and Display}. On an LCD panel, a single red, green, or blue element is referred to as a subpixel. The light emitted by these subpixels is refracted by the lens, deflecting into different spatial regions to form viewing zones. The angle of deflection is related to the distance of the subpixel center from the optical axis of the lens unit. Specifically, subpixels on the LCD panel that are equidistant from the optical axis center of the lens grating unit should display information from the same viewpoint image to ensure a correct viewing experience. For a lenticular lens grating, there are three primary hardware parameters:
\begin{itemize}
    \item Tilt angle $\alpha$: The angle by which the grating is inclined from the vertical direction. To avoid moiré patterns resulting from the periodicity of the grating and the subpixel arrangement, the stripes of the lenticular 3D display's grating units are inclined at a certain angle from the vertical direction \cite{van1999image}. A clockwise rotation from the vertical direction is considered positive, and a counterclockwise rotation is negative.
    \item Line count $L_{\rm x}$: The width of the subpixel area covered by the grating unit in the horizontal direction. As shown in Fig. \ref{fig:display_sub2}, the height of the RGB subpixels on the LCD panel is three times their width. A subpixel unit's width is defined as a unit length of 1, and its height is three times the unit length.
    \item Offset $K_{\rm offset}$: The horizontal shift of the lenticular lens array relative to the LCD. Due to manufacturing discrepancies, there is often a misalignment between the grating and the LCD panel.
\end{itemize}
Interlacing is a two-step process. The first step involves calculating the number of viewpoints corresponding to each subpixel on the LCD panel. The second step is sampling the color values of the subpixels from the respective viewpoint images. For any subpixel on the LCD panel with coordinates ($x, y, k$), the formula to calculate the distance from the left edge of this subpixel to the left edge of its corresponding lenticular lens grating unit is as follows \cite{van1999image, shen2023virtual}:
\begin{equation}
    \label{eq:doffset}
    d_{\rm offset} = 3 \cdot y + 3 \cdot x \cdot \tan(\alpha) + k - K_{\rm offset}
\end{equation}
\begin{equation}
    \label{eq:xoffset}
    x_{\rm offset} = d_{\rm offset} \mod L_{\rm x}
\end{equation}
where ($x, y, k$) represents the subpixel at the $x$-th row, $y$-th column, and $k$-th color channel, with the top-left subpixel having coordinates (0,0,0). $d_{\rm offset}$ denotes the horizontal distance from the left edge of the subpixel to the left edge of the most distant grating unit. $x_{\rm offset}$ represents the distance from the left edge of the subpixel to the left edge of the corresponding grating unit. $x_{\rm offset}$ uniquely determines the number of viewpoints for that subpixel and is confined within the interval $\left[0, L_{\rm x}\right)$. When $0 \leq x_{\rm offset} < \frac{L_{\rm x}}{N_{\rm v}}$, the subpixel's viewpoint number is 1; when $\frac{L_{\rm x}}{N_{\rm v}} \leq x_{\rm offset} < \frac{L_{\rm x}}{N_{\rm v}} \times 2$, the viewpoint number is 2, and so on. Therefore, the viewpoint number $v$ for a subpixel with a distance $x_{\rm offset}$ from the left edge of its grating unit is given by:
\begin{equation}
    \label{eq:v}
    v = \lfloor N_{\rm v} \cdot \frac{x_{\rm offset}}{L_{\rm x}} \rfloor
\end{equation}
where the floor function $\lfloor \cdot \rfloor$ indicates taking the integer part. The matrix of all subpixel viewpoint numbers is known as the viewpoint number matrix $M_{\rm v}^{h \times w \times 3}$, where $h$ and $w$ correspond to the height and width of the LCD panel, respectively. $M_{\rm v}$ is uniquely determined by the hardware parameters. Once $M_{\rm v}$ is computed, the images from each viewpoint are scaled to match the size of the encoded image, and subpixels from the same position in the corresponding viewpoint images are extracted and filled into the encoded image according to $M_{\rm v}$. The interlacing process is summarized in Alg. \ref{alg:interlace}. To ensure an accurate mapping relationship, the encoded image's width and height must match the LCD panel's dimensions exactly, and it must be displayed in full screen.

\begin{algorithm}
		\caption{Interlacing\\
			Width $w$ and Height $h$ of the LCD panel \\
			Tilt angle $\alpha$, Line count $L_{\rm x}$, Offset $K_{\rm offset}$\\
            Multi-View Image Matrix $M_{\rm m}^{N_{\rm v} \times \hat{h} \times \hat{w} \times 3}$\\
            Encoded Image Matrix $M_{\rm e}^{h \times w \times 3}$, Viewpoint Matrix $V^{h \times (w \times 3)}$}
		\label{alg:interlace}
		\begin{algorithmic}

            \State $M_{\rm m}^{N_{\rm v} \times h \times w \times 3} \gets$ Resize($M_{\rm m}^{N_{\rm v} \times \hat{h} \times \hat{w} \times 3}$, $w$, $h$) \Comment{Upsample}
			\ForAll{$x$ $\textbf{in}$ range($h$)}
			\ForAll{$y$ $\textbf{in}$ range($w$)}
            \ForAll{$k$ in range($3$)} \Comment{Iterate over RGB}
			
			\State $v \gets$ CalculateView($\alpha$,$L_{\rm x}$,$K_{\rm offset}$,$x$,$y$,$k$) \Comment{Eq.  \ref{eq:doffset}, \ref{eq:xoffset}, \ref{eq:v}}
            \State $V\left[x, 3y + k\right] \gets v$\Comment{Align with LCD panel}
			\State $M_{\rm e}\left[ x, y, k\right] \gets M_{\rm m}[v, x, y, k]$ \label{eq:sample}
			
			\EndFor
			\EndFor
            \EndFor
            
	\Return $V, M_{\rm e}$
			
		\end{algorithmic}
	\end{algorithm}

As shown in Alg. \ref{alg:interlace} at line \ref{eq:sample}, the color value at location ($x$, $y$, $k$) in the encoded image matrix $M_{\rm e}$ is directly sampled from the corresponding viewpoint image matrix at the same position. While a common alternative involves a weighted average of neighboring viewpoint color values, this technique, once beneficial for LFDs with sparse viewpoints, is now obsolete. Modern LFDs, with their increased viewpoint density and reduced crosstalk, would suffer from significant crosstalk artifacts using this method \cite{woods2012crosstalk, shen2023virtual}, thus we do not consider this approach.

Due to high computational complexity, the resolution of multi-view images is generally significantly lower than that of the encoded image \cite{guan2020parallel, fink2023efficient}. Extensive interpolation operations can severely impact the display depth, generating blurred artifacts that adversely affect the 3D display quality \cite{nam2017flat, he2024assessment}. In practice, utilizing texture look-up with OpenGL can circumvent global interpolation, thereby reducing the interlacing time. However, it is noteworthy that interlacing can be completed in an extremely short time with GPU acceleration. The majority of the time consumption remains in rendering the multi-view images. The primary approach to reduce the amount of computation is to reduce the rendering resolution for each viewpoint. Yet, the upsampling operation during interlacing, after reducing the viewpoint resolution, can still result in display blurring. Therefore, to achieve the best display effects, it is necessary for us to seek an alternative rendering paradigm.

\subsection{Neural Radiance Fields}
NeRFs (Neural Radiance Fields) \cite{mildenhall2021nerf} learn the mapping from every 3D spatial position $\bm x = \left( x, y, z \right)$ and viewing direction $\bm d = \left( \theta, \phi \right)$ to the volumetric density $\sigma$ and color $\bm c = \left( r, g, b \right)$ at that point, which can be formalized as:
\begin{equation}
    \label{eq:nerfs}
    \left( \sigma, \bm c \right) = F_{\Theta} \left( \bm x, \bm d \right)
\end{equation}
Subsequently, the color along a ray is obtained by sampling points along the ray and performing volumetric rendering using these points' volumetric densities $\{\sigma_{i}\}$ and colors $\{ \bm c_{i} \}$ \cite{max1995optical}, given by:
\begin{equation}
    \label{eq:alpha_blending}
    \bm C = \sum_{i} \alpha_i \cdot T_i \cdot \bm c_i
\end{equation}
\begin{equation}
    \alpha_i = 1 - e^{-\sigma_i\delta_i}, \quad T_i = \prod_{j=1}^{i-1}(1 - \alpha_j)
\end{equation}
where $\alpha_i$ represents the alpha value at sampling point $i$, $T_i$ is the transmittance at that point, and $\delta_i$ denotes the distance between samples. The original NeRF \cite{mildenhall2021nerf} employs a view-dependent Multi-Layer Perceptron (MLP) to model the radiance field. Ray-casting inherently allows NeRFs to perform rendering in ray-order.
\begin{figure*}[!t]
    \centering
    \includegraphics[width=\textwidth]{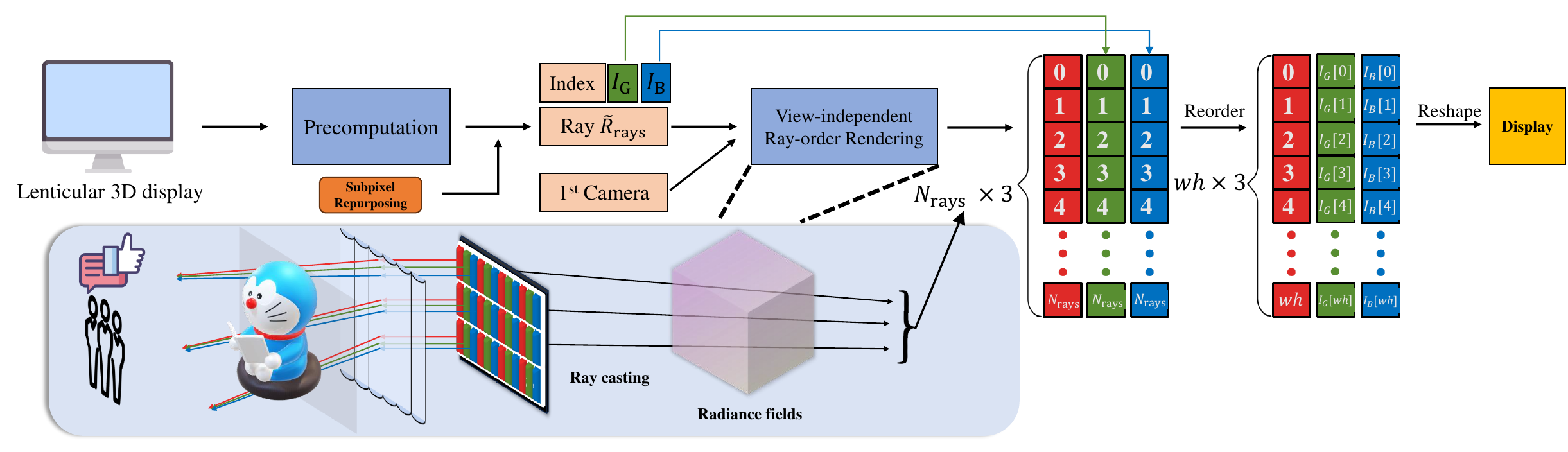}
    \caption{Overview of the DirectL. Initially, ray configurations and the requisite index for rearrangement are calculated offline based on the hardware parameters of any given lenticular 3D display. Subsequently, the origin $\bm o_i$ and direction $\bm d_i$ of each ray are determined according to ray configurations and the current center of the light field. Following this, view-independent ray-order rendering of the radiance field is conducted to ascertain the color of each ray. Finally, the G and B channels are reordered to produce a light field image ready for direct display.}
     \label{fig:viewind}
\end{figure*}

\subsection{Gaussian Splatting Radiance Field}
Unlike neural network-based implicit radiance fields, the Gaussian Splatting Radiance Field (3DGS) \cite{kerbl20233d} is an explicit radiance field that represents a scene as a collection of 3D anisotropic Gaussian ellipsoids. Specifically, each 3D Gaussian ellipsoid is parameterized by its mean $\bm \mu_i$, covariance $\bm \Sigma_i$, opacity $o_i$, and view-independent color $\bm c_i$ parameterized by spherical harmonics. By default, the model's coordinate center is at the origin of the coordinate system, and the Gaussians are defined as follows:
\begin{equation}
\label{eq:gaussian}
    G(\bm x) = e^{-\frac{1}{2}(\bm x - \bm \mu_i)^\top\bm \Sigma_i^{-1}(\bm x - \bm \mu_i)}
\end{equation}
To maintain the semi-definiteness of $\bm \Sigma$ during optimization, $\bm \Sigma$ is represented using a scaling matrix $\bm S$ and a rotation matrix $\bm R$:
\begin{equation}
    \bm \Sigma = \bm R \bm S \bm S^\top \bm R^\top
\end{equation}
3DGS employs tile-based splatting rasterization to accelerate rendering. Specifically, the Gaussians within the view frustum are first projected onto a 2D plane. Given the viewing transformation $\bm W$, the projected 2D covariance matrix $\bm \Sigma'$ is computed using \cite{zwicker2001ewa}:
\begin{equation}
    \bm \Sigma' = \bm J \bm W \bm \Sigma \bm W^\top \bm J^\top
\end{equation}
where $\bm J$ is the Jacobian of the affine approximation of the projective transformation. The image is then divided into $16 \times 16$ tiles, and the projected Gaussians intersecting with each tile are sorted by depth. For every pixel within a tile, its color is computed by alpha compositing the opacity and color of all the Gaussians covering that pixel in depth order, following the same computation method as in Eq. \ref{eq:alpha_blending}, but applied cumulatively to all Gaussians covering the pixel. The $\alpha_i'$ in Eq. \ref{eq:alpha_blending} is obtained by multiplying the Gaussian's opacity and the power of the projected 2D Gaussian at the pixel position, which can be defined as follows:
\begin{equation}
    \label{eq:g_a}
    \alpha_i = o_i \cdot e^{-\frac{1}{2}(\bm x' - \bm \mu_i')^\top\bm \Sigma_i'^{-1}(\bm x' - \bm \mu_i')}
\end{equation}
where $\bm x'$ and $\bm \mu_i'$ are the projected coordinates. In practice, the rendering of 3DGS is highly parallelized, where tiles and pixels correspond to blocks and threads in CUDA, respectively. Each tile's pixels access a segment of shared memory for efficient alpha blending.

\section{Overview}
DirectL is divided into three steps (see Fig. \ref{fig:viewind}): (1) Calculate the ray configurations post subpixel repurposing and the index matrix required for rearrangement, based on hardware parameters (offline, see Sec. \ref{Subpixel Repurposing}).  (2) Perform view-independent rendering in ray-order (online, see Sec. \ref{View-Independent Ray-Casting for NeRFs} and \ref{3DGS}). (3) Reorder the G and B color channels according to the index matrix (online).
Specifically, we precisely compute offline the transformation matrix for each required ray relative to the center of the frontal light field, based on the hardware parameters of the light field screen. The order of rays is arranged to correspond with the source rays of the R channel in the encoded image matrix. Once the ray order is fixed, the index for the rearrangement of the G and B channels into the encoded image is computed offline. Subsequently, for both NeRFs and 3DGS, we execute view-independent rendering on a per-ray basis. Finally, the G and B channels of the rendered result are reordered using the previously calculated indices to obtain the encoded image for 3D display.
\begin{figure}[!t]
    \centering
    \includegraphics[width=0.5\textwidth]{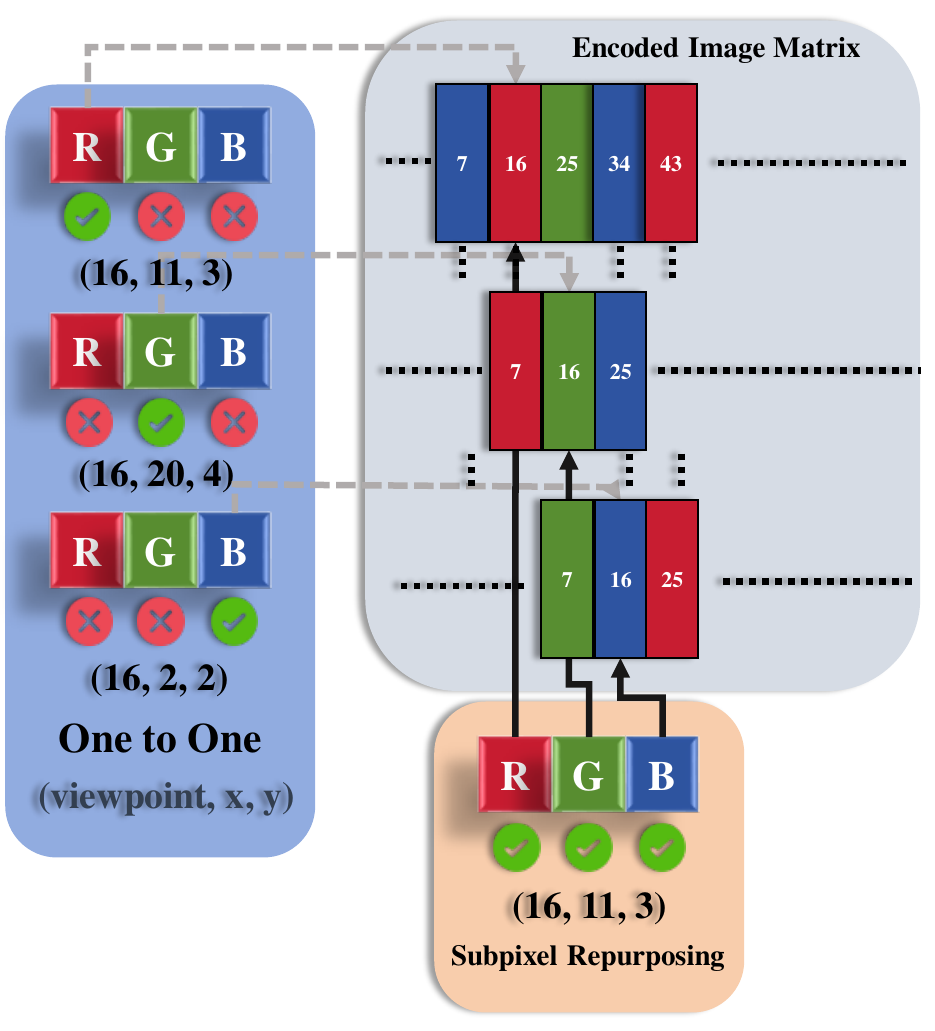}
    \caption{Our subpixel repurposing scheme. In the standard paradigm, the one-to-one mapping approach only leverages a single subpixel from the multi-viewpoint image pixels, with the remaining subpixels being discarded. Following the display principle, we repurpose these discarded subpixels during the interlacing process, thereby reducing the number of pixels that require rendering. }
    \label{fig:sr}
\end{figure}
\section{Subpixel Repurposing}
\label{Subpixel Repurposing}
By leveraging a trained radiance field to render light field images, it is also feasible to render multi-view images followed by interlacing. However, in contrast to the conventional Multiview Rendering (MVR) approach \cite{halle1998multiple, unterguggenberger2020fast}, we introduce the Ray-Order Rendering (ROR) algorithm following the \cite{nam2017flat}. ROR represents a generalized rendering strategy for light field displays that is not limited by the number of viewpoints. It directly renders in the ray-order required for light field displays without interpolation, which is more aligned with the principles of light field display. Since each ray can be precisely controlled, this approach can minimize the blur effect and maximize visual clarity and depth perception \cite{nam2017flat, he2024assessment}. The question now arises as to how these rays can be determined?

According to Alg. \ref{alg:interlace}, each subpixel in the encoded image matrix $M_{\rm e}$ is defined by the unique indices ($v$, $x^{\rm v}$, $y^{\rm v}$, $k$), with $x^{\rm v}$ and $y^{\rm v}$ indicating the subpixel's position within the image of viewpoint $v$. The subpixel corresponds to the spatial light ray that connects to the pixel at \((x^{\rm v}, y^{\rm v})\) from the camera at viewpoint \(v\). Typically, \(x^{\rm v}\) and \(y^{\rm v}\) reflect the subpixel's internal coordinates within the encoded image, and color channel \(k\) maintains consistency. Given positional dependencies, if the red (R) channel subpixel at \((x^{\rm v}, y^{\rm v})\) for viewpoint \(v\) is sampled, the subpixels of the green (G) and blue (B) channels at the identical location remain unsampled. This occurs due to the distinct $x_{\rm offset}$ for the triplet of subpixels beneath a singular pixel in the encoded image, guaranteeing separate originating viewpoints. 

Under the standard paradigm, the requisite light rays for display equate to the LCD panel's subpixel count—triple that of the pixels. Rendering solely one color channel in the existing pipeline for radiance fields fails to boost performance, thereby almost tripling the computational burden. For typical 2D LCD displays, RGB subpixels within the same pixel unit blend at varying intensities to produce diverse colors as perceived by the human eye. For lenticular 3D displays, LCD panel subpixels diverge through the lenticular lenses into distinct spatial zones. Though the light path is theoretically reversible, grating production constraints preclude precise RGB subpixel color blending determinations. Consequently, the standard interleaving method employs a simplistic equal-position sampling approach, yielding incomplete pixel combinations in the encoded image and slight color inaccuracies to a certain extent. Despite the precise light path being indeterminate, RGB subpixels from different viewpoints do not mix, and neither do those from the same viewpoint that are far apart. The general rule is that the red (R) subpixel under one grating unit converges into the eye along with the closest green (G) and blue (B) subpixels from the same viewpoint within the same grating unit \cite{yeh2009optics, zhang2017illumination}. Inspired by this, we propose repurposing idle subpixels to appropriate positions within the encoded image, as shown in Fig. \ref{fig:sr}. Given the intractability of the exact light path and the difficulty of achieving a global optimum, we employ a heuristic approach.

The subpixel repurposing generally includes four steps. 
First, we calculate the viewpoint matrix $V^{h \times (w \times 3)}$, containing the view index for each subpixel, using Alg. \ref{alg:interlace}. This calculation is constant and based solely on the lenticular lens's hardware parameters.

Secondly, in the encoded image, if the subpixels R, G and B of the same pixel from the same viewpoint are too far from each other, the imaging quality will be degraded. To resolve this issue, we partition the LCD panel into multiple non-intersecting areas, each comprising $P_{\rm w}$ grating units, with subpixel repurposing carried out individually within each area.

The third phase focuses on documenting the origination information of subpixels in the encoded image, tracing their exact positions within the viewpoint image and encapsulating these data within the encoded image index matrix $M_{\rm i}$. This matrix $M_{\rm i}^{h \times w \times 3 \times 4}$ is structured to map one-to-one with every position in the encoded image, amassing $h \times w \times 3$ subpixel information entries spanning four dimensions: the viewpoint, horizontal and vertical coordinates, and color channel number. Additionally, the color buffer $B_{\rm rgb}$ is employed to systematically archive idle subpixel information for each viewpoint image, including coordinates and the channel number.

To determine the subpixel that should be placed at position $(x, y, k)$ in the encoded image, we need to check if there are idle subpixels of channel $k$ from the same viewpoint $v$ ($v \gets V(x, 3 \times y + k)$) that can be repurposed. If there is no such subpixel in $B_{\rm rgb}$, we add the current position $(x, y)$ and channel number $k$ to $M_{\rm i}(x, y, k)$:
\begin{equation}\label{eq:resubpixl_or}
M_{\rm i}(x, y, k) = (v, x, y, k)
\end{equation}
which means we sample the subpixel from the same position of the $v$-th viewpoint image, and store the remaining idle subpixel information at position $(x, y)$ in the same viewpoint image into $B_{\rm rgb}$:
\begin{equation}\label{eq:resubpixl_cache}
    \begin{aligned}
    B_{\rm rgb}(v, (k - 1)\mod 3) &= (x, y) \\
    B_{\rm rgb}(v, (k - 2)\mod 3) &= (x, y) \\
    \end{aligned}
\end{equation}
However, if there is a idle subpixel of channel k from the viewpoint $v$ in the $B_{\rm rgb}$, retrieve the coordinates $(x', y')$ of the available subpixel from the buffer and assign them to $M_{\rm i}(x, y, k)$.
\begin{equation}\label{eq:resubpixl_rs}
    \begin{aligned}
        ({x}', {y}') &= B_{\rm rgb}(v, k) \\
        M_{\rm i}(x, y, k) &= (v, {x}', {y}', k)
    \end{aligned}
\end{equation}
Concurrently, deallocate the position $({x}', {y}')$ from the buffer $B_{\rm rgb}(v, k)$.

The final stage's goal is to pinpoint the next position suitable for repurposing subpixels within a consistent viewpoint. A pivotal insight underpins this procedure: Examination of Alg. \ref{alg:interlace} shows that, within the resultant encoded image $M_{\rm e}$, subpixels tied to a singular viewpoint are organized in a fashion nearly parallel to the grating units. Importantly, the R, G, and B subpixels adjacent to this linear alignment adhere meticulously to the imaging standards of the corresponding viewpoint. This arrangement points to an ideal trajectory for the subpixel repurposing, facilitating the accurate reconstruction of the original image's viewpoint data. However, as subpixels of the same viewpoint are not strictly arranged linearly, a minimum search range is considered. We propose a general rule for determining the area based on Alg. \ref{alg:interlace}:
\begin{equation}\label{eq:resubpixl_cr}
    \begin{aligned}
    x_{\rm low} &= \min (x + 2, h - 1)\\
    y_{\rm low} &= \max (y - N_{\rm v} / ( 3 \times L_{\rm x})), 0)\\
    x_{\rm high} &= \min(x + 2 \times (N_{\rm v} / L_{\rm x}) + N_{\rm v}/ (2 \times L_{\rm x}), h - 1)\\
    y_{\rm high} &= \max (y, y_{\rm low})
    \end{aligned}
\end{equation}
If no suitable candidate position is found, it is necessary to clear the buffer for the current viewpoint and switch to another to proceed with the repurposing algorithm.

By repeating stages three and four until $M_{\rm i}$ is fully populated, we gather the index of the subpixel at each location in the encoded image. This thorough collection enables us to assemble the desired encoded image using $M_{\rm i}$. The pseudocode for the subpixel repurposing is shown in Alg. \ref{alg:resubpixl}.

\begin{algorithm}[!t]
		\caption{Subpixel repurposing\\
			Width $w$ and Height $h$ of the LCD panel \\
			Viewpoint number $N_{\rm v}$, Line count $L_{\rm x}$, Grating unit number $P_{\rm w}$\\
            Encoded Image Index Matrix $M_{\rm i}^{h \times w \times 3 \times 4}$
            }
		\label{alg:resubpixl}
		\begin{algorithmic}

            \State $V^{h \times (w \times 3)} \gets$ Get the Viewpoint Matrix from Alg. \ref{alg:interlace}.
            
            \State $I_{\rm areas}$ $\gets$ SplitArea($h$, $w$, $P_{\rm w}$) 

            \State $B_{\rm rgb} \gets$ CreateSubpixelBuffer($N_{\rm v}$, $I_{\rm areas}$)
            
            \ForAll{subpixels($x_{\rm i}, y_{\rm i}, k_{\rm i}, I_{\rm area}$) \textbf{in} $I_{\rm areas}$}
            \State $v_{\rm i} \gets$ GetViewpoint($V$, $x_{\rm i}$, $y_{\rm i}$, $k_{\rm i}$)
            


            \While{$ \: M_{\rm i}(x_{\rm i}, y_{\rm i}, k_{\rm i})$ is empty \textbf{and} $(x_{\rm i}, y_{\rm i}, k_{\rm i})$ \textbf{in} $I_{\rm area}$} 
                
                \If{$B_{\rm rgb}(v_{\rm i}, k_{\rm i})$ is not None}
                    \State RepurposeSubpixel($M_{\rm i}, x_{\rm i}, y_{\rm i}, k_{\rm i}, v_{\rm i}, B_{\rm rgb}$) \Comment{Eq.\ref{eq:resubpixl_rs}}
                    \State ClearUsedSubpixel($x_{\rm i}, y_{\rm i}, k_{\rm i}, v_{\rm i}, B_{\rm rgb}$)
                \Else
                    \State SampleNewSubpixel($M_{\rm i}, v_{\rm i}, x_{\rm i}, y_{\rm i}, k_{\rm i}$) \Comment{Eq.\ref{eq:resubpixl_or}}
                    \State AddUnusedSubpixel($x_{\rm i}, y_{\rm i}, k_{\rm i}, v_{\rm i}, B_{\rm rgb}$) \Comment{Eq.\ref{eq:resubpixl_cache}}
                \EndIf 
                \State $x_{\rm low}, y_{\rm low}, x_{\rm high}, y_{\rm high}\gets$ CalculateSearchRange($x_{\rm i}$, $y_{\rm i}$, $h$, $w$, $N_{\rm v}$, $L_{\rm x}$) \Comment{Eq.\ref{eq:resubpixl_cr}}
                
                \State {$x_{\rm i}, y_{\rm i}, k_{\rm i}, s \gets$ SearchTheSameViewpoint($V$, $x_{\rm low}$, $x_{\rm high}$, $y_{\rm low}$, $y_{\rm high}$, $k_{\rm i}$, $v_{\rm i}$)}

                \If{$s$ is False}  
                    \State {Break}  \Comment{No position from the $v_{\rm i}$ viewpoint}
                \EndIf

            \EndWhile 

            \State ClearSubpixelBuffer($B_{\rm rgb}$, $v_{\rm i}$) \Comment{Clear the buffer of viewpoint $v_{\rm i}$}
			\EndFor 
            
	\Return $M_{\rm i}$
			
		\end{algorithmic}
	\end{algorithm}

We sequentially number all subpixels of the encoded image in row-major order as $\{1, 2, \ldots, 3wh\}$, where the red (R) subpixel at the top-left corner is numbered 1, and the blue (B) subpixel at the bottom-right corner is numbered $3wh$. We then segregate the subpixel indices for the three color channels: $R_{\rm I} = \{1, 4, \ldots, 3wh-2\}$, $G_{\rm I} = \{2, 5, \ldots, 3wh-1\}$, and $B_{\rm I} = \{3, 6, \ldots, 3wh\}$. Following subpixel repurposing, we obtain a set of unique indices $U = \{(v_i, x_i^{\rm v}, y_i^{\rm v}, k_i), i \in R_{\rm I} \cup G_{\rm I} \cup B_{\rm I}\}$ for all subpixels in the encoded image.
We define ordered, non-repetitive sets as follows, where the subscript $i$ represents the sequence of elements within the set:
\begin{align}
    U_{\rm rays} &= \{ (v_i, x_{i}^{\rm v}, y_{i}^{\rm v}), i \in U_{\rm I}\}, R_{\rm rays} = \{ (v_i, x_{i}^{\rm v}, y_{i}^{\rm v}), i \in R_{\rm I}\} \\
    G_{\rm rays} &= \{ (v_i, x_{i}^{\rm v}, y_{i}^{\rm v}), i \in G_{\rm I}\}, B_{\rm rays} = \{ (v_i, x_{i}^{\rm v}, y_{i}^{\rm v}), i \in B_{\rm I}\}
\end{align}
where $U_{\rm rays}$ is the unique set of tuples $(v_i, x_i^{\rm v}, y_i^{\rm v})$ derived from the first three elements of each tuple in $U$, and $U_{\rm I}$ represents the index set for $U_{\rm rays}$. The cardinality $|U_{\rm rays}| = N_{\rm rays}$ corresponds to the number of rays, which is the number of pixels we need to render for one frame on the light field display. It is evident that $1 \leq \frac{N_{\rm rays}}{wh} \leq 3$, and $R_{\rm rays}, G_{\rm rays}, B_{\rm rays} \subseteq U_{\rm rays}$.
We then perform three separate re-orderings of the elements in $U_{\rm rays}$ to align the sequence of the first $wh$ elements with those in $R_{\rm rays}$, $G_{\rm rays}$, and $B_{\rm rays}$, resulting in matrices $\widetilde{R}_{\rm rays}$, $\widetilde{G}_{\rm rays}$, and $\widetilde{B}_{\rm rays}$, each with the shape ($N_{\rm rays}, 3$), differing only in the order of their rows.
Subsequently, we calculate the indices of each row in $\widetilde{R}_{\rm rays}$ within $\widetilde{G}_{\rm rays}$ and $\widetilde{B}_{\rm rays}$, forming index matrices $I_{\rm G}$ and $I_{\rm B}$. By rendering the colors in the order of $\widetilde{R}_{\rm rays}$ and then reordering the G and B channels according to the indices $I_{\rm G}$ and $I_{\rm B}$, we obtain the encoded image by taking the first $wh$ rows and reshaping them into the format ($h$, $w$, 3). For each light field display panel, we precompute $\widetilde{R}_{\rm rays}$ and the index matrices $I_{\rm G}$, $I_{\rm B}$.

\section{View-Independent Ray-Casting for NeRFs}
\label{View-Independent Ray-Casting for NeRFs}
Given the extrinsic matrix of light field center, it is straightforward to compute the extrinsic matrices for cameras at all viewpoints as follows:
\begin{equation}
    T^{v \times 4 \times 4} = R^{v \times 4 \times 4} \cdot L_{0}^{4 \times 4}
\end{equation}
where $L_0$ represents the extrinsic matrix of light field center, $R$ is the rotation matrix that transforms the light field center to all viewpoint cameras, and $T$ denotes the corresponding extrinsic matrices for all viewpoint cameras. 

The origin $\bm o_i$ and direction $\bm d_i$ for each ray can be computed based on $\widetilde{R}_{\rm rays}$, 

\begin{equation}
    v_i, x_i^{\rm v}, y_i^{\rm v} = \widetilde{R}_{\rm rays}\left[i\right]
\end{equation}
\begin{equation}
    \bm o_i = (T \left[ v_i, 0, 3\right],T \left[ v_i, 1, 3\right],T \left[ v_i, 2, 3\right], 1)
\end{equation}
   \begin{equation}
       \bm d_i = T^{4 \times 4}_{[v_i]} \cdot \begin{bmatrix}\frac{x_i^{\rm v}-\frac{h}{2}}{f}\\\frac{y_i^{\rm v}-\frac{w}{2}}{f}\\-1\\0\end{bmatrix}_{4 \times 1}, i=0,...,N_{\bm rays}-1
\end{equation}

Since NeRFs\cite{mildenhall2021nerf, muller2022instant, hedman2021baking} naturally support per-pixel rendering, we propose a straightforward yet effective strategy to replace all rays cast through a single camera with the previously calculated rays. Clarifying a key distinction, while NeRFs' color computation is view-dependent, rendering the light field image using the $\widetilde{R}_{\rm rays}$ set is view-independent, as it pertains to a discrete frontal light field determined by hardware parameters alone. Subpixel repurposing optimizes color allocation, reducing the ray count for each frame without changing their directions. Following the acquisition of all ray colors $C^{N_{\rm rays}\times3}$, we reorder the G and B channels based on $I_{\rm G}$ and $I_{\rm B}$. The first $wh$ rows (the number of pixels on the LCD panel) are reshaped into ($h, w, 3$), resulting in a light field image frame.

We highlight that the DirectL paradigm for rendering light field images is not exclusive to NeRFs. In fact, any method that inherently supports image-order rendering can readily adopt the DirectL framework.  A case in point is the Neural Light Fields (NeLFs) \cite{attal2022learning, cao2023real, wang2022r2l}, which directly learning the mapping between rays and colors, circumventing the need for ray marching. As these methods also render in image-order, they are equally amenable to application in light field display technology using the aforementioned methodology. 

\section{Ray-Order 3DGS}
\label{3DGS}
For light field displays, we aim to leverage the high-quality rapid reconstruction and active research momentum of 3DGS. However, current 3DGS-based works use the original 3DGS tile-based rasterizer, which follows an object-order rendering approach. Our analysis underscores the importance of per-ray rendering for improving light field rendering efficiency. The MVR method incurs computational costs proportional to the number of viewpoints \(N_{\rm v}\). Even without considering memory usage, parallel rendering of multiple viewpoints to maximize GPU utilization is significantly slower than rendering a single viewpoint of equivalent resolution \cite{shen2023virtual, fink2023efficient}. For light field displays, this redundancy necessitates the design of ray-order 3DGS, which also provides substantial tolerance in optimizing rendering speed.

The original 3D Gaussian Splatting (3DGS) tile-based rasterization process consists of four steps: 1) Projecting 3D Gaussian ellipsoids onto a 2D plane, 2) Calculating which tiles are covered by the projected Gaussians (approximated as circles), 3) Determining the order of Gaussians within each tile, and 4) Computing the color of each pixel through alpha-blending. Sorting and alpha-blending constitute the majority of the computational workload, with tile partitioning being key to reducing redundant calculations. A straightforward idea is to consider whether we can compute the pixel colors only for the locations required by $\widetilde{R}_{\rm rays}$. Since the pixels to be rendered are irregularly but uniformly distributed across each viewpoint, efficient computation of intersections with Gaussians for these pixels is not feasible through tile partitioning. Using the original square tile division would save only the computational effort of pixel color calculations for unnecessary locations. However, as each pixel in a tile is a thread, the overall speed would not significantly improve if any pixel within a tile requires computation. Based on this analysis, we propose to implement an efficient ray-order rendering for 3DGS by utilizing ray casting instead of rasterization. 

Our ray-order 3DGS, based on ray casting, is streamlined into three steps: 1) Identifying intersections between each ray and the Gaussians, 2) Calculating and sorting the intersection points by distance, and 3) Determining the color of each ray through alpha-blending. Leveraging efficient algorithms from ray tracing \cite{meister2021survey}, we utilize a Bounding Volume Hierarchy (BVH) for efficient ray traversal. Specifically, we construct a compact Surface Area Heuristic (SAH)-based 8-wide BVH from 3D Gaussians, following \cite{ylitie2017efficient}, where each leaf node represents the axis-aligned bounding box (AABB) of the Gaussians. Minimizing the SAH cost during BVH construction enhances the performance of ray intersection tests. The cost is defined as:
\begin{equation}
    SAH = \sum_{n \in I} A_n \cdot c_{\rm node} + \sum_{n \in L} A_n \cdot P_n \cdot c_{\rm prim}
\end{equation}
where \( I \) and \( L \) represent the sets of internal and leaf nodes of the BVH, respectively. \( A_n \) is the surface area of node \( n \)'s AABB relative to the root node's AABB. \( P_n \) is the number of primitives within leaf node \( n \), and \( c_{\rm node} \) and \( c_{\rm prim} \) are constants representing the computational costs for ray-node and ray-Gaussian intersection tests, respectively. Unlike ray-mesh intersection tests that find the nearest primitive, ray-Gaussian intersection tests must identify all Gaussians affecting the ray's transmittance. Following \cite{keselman2022approximate}, for a ray with origin \( \bm o_i \) and direction \( \bm d_i \), we find the intersection point of highest probability by maximizing \( (\bm o_i + t_j\bm d_i - \bm \mu_j)^\top\bm \Sigma_j^{-1}(\bm o_i + t_j\bm d_i - \bm \mu_j) \), yielding a linear solution: 
\begin{equation} \label{eq:ray_intersection}
    t_j = \frac{(\bm \mu_j - \bm o_i)^\top\bm\Sigma_j^{-1}\bm d_i}{\bm d_i^\top\bm\Sigma_j^{-1}\bm d_i} 
\end{equation}
Using Eq. \ref{eq:gaussian}, we compute \( \alpha_j \) in Eq. \ref{eq:g_a} as:
\begin{equation} \label{eq:ray_alpha}
    \alpha_j = o_jG(\bm o_i + t_j \bm d_i) 
\end{equation}
Each ray maintains a fixed-size array to implement a max-heap for intersection distances and stores the indices of intersecting Gaussians. Upon finding an intersecting Gaussian, the intersection point \( t_j \) is calculated. If the heap is not full, \( t_j \) is inserted; if full, \( t_j \) is compared with the root node (the farthest distance), and the root is replaced if \( t_j \) is smaller. After ray traversal, the heaps for each ray are used to sort the Gaussians, followed by alpha-blending for color computation. We note that the heap's capacity $c_{\rm h}$ directly impacts image quality and rendering efficiency, presenting a trade-off (see Sec. \ref{abl}). A high-level per-ray overview of ray-order 3DGS is summarized in Alg. \ref{alg:ray-order-3dgs}. 

\begin{algorithm}
	\caption{Ray-order 3DGS\\
		3DGS's BVH tree $bvh$ \\
        Ray origin $o$, Ray direction $d$\\ 
        max-heap $h$ 
		}
	\label{alg:ray-order-3dgs}
	\begin{algorithmic}
		\ForAll {$g$ \textbf{in} IntersectsBVH($bvh$, $o$, $d$)}
            \State $t$,$alpha\gets$  CalculateIntersectionAndAlpha($o$,$d$,$g$) \Comment{Eq.  \ref{eq:ray_intersection} \ref{eq:ray_alpha}} 
		\State  $rgb\gets$ CalculateFromSHS($d$, $g$) 
            \If{Full($h$)}
                \If{$t$ < Getmax($h$)}
                    \State Popmax($h$) 
                    \State AddToHeap($h$,$t$,$alpha$,$rgb$)   
                \EndIf
            \Else
                 \State  AddToHeap($h$,$t$,$alpha$,$rgb$) 
            \EndIf
            \EndFor
            
        \Return AlphaBlending(Sort($h$))
	\end{algorithmic}
\end{algorithm}

While the aforementioned scheme enables a relatively efficient ray-order 3DGS, we candidly acknowledge its limitations: 1) The use of rasterization in the optimization process introduces a degree of misalignment, leading to some degradation (see Sec. \ref{Baseline Comparisons}). 2) The lack of tile division results in slower rendering speeds for a single viewpoint image of the same resolution compared to rasterization methods. We propose potential solutions to these issues for future work. For the first limitation, refining the trained 3DGS with differentiable ray casting, fine-tuning only the opacity and spherical harmonics while retaining the Gaussian means and covariances, can reduce training-inference inconsistencies without requiring BVH modifications during training. For the second, the primary computational expense in ray-order 3DGS rendering lies in ray-intersection and sorting. The ray-intersection computation could be accelerated using a parameterized 2D Gaussian tangent plane \cite{huang20242d}. Order-Independent Transparency (OIT) techniques \cite{keselman2022approximate, mcguire2013weighted} could eliminate the need for sorting. Additionally, deformable 3DGS \cite{wu20234d, yang2023deformable} requires dynamic adjustment of the BVH \cite{wald2007ray, wald2007fast, karras2013fast} to accommodate changes in the scene's geometry. 

\begin{table}[!t]
\caption{Calibrated light field display parameters.}
\label{tab:lfds}
\resizebox{0.6\columnwidth}{!}{%
\begin{tabular}{llll}
\hline
Parameter  & 7.9-inch       & 15.6-inch     & 65-inch       \\ \hline
Width $w$& 1536 px        & 3840 px       & 7680 px       \\
Height $h$& 2048 px        & 2160 px       & 4320 px       \\
Line count $L_{\rm x}$& 6.2221 px      & 5.3344 px     & 9.3597 px     \\
Tilt angle $\alpha$& 10.8232$^\circ$& 6.8526$^\circ$& 8.6517$^\circ$\\
Offset $K_{\rm offset}$& 4.2077 px      & 1.2547 px     & 23.6677 px    \\
FOV $\theta_{\rm d}$& 40$^\circ$& 53$^\circ$& 80$^\circ$\\
Viewpoints $N_{\rm v}$& 48             & 60            & 96            \\ 
LR & 420$\times$560 px & 800$\times$450 px & 1600$\times$900 px \\
MR & 768$\times$1024 px & 1920$\times$1080 px & 3840$\times$2160 px \\ \hline
\end{tabular}%
}
\end{table}

\begin{table*}[!t]
\caption{Quantitative comparison of our method with the standard paradigm leveraging two different NeRFs methods (Instant-NGP, SNeRG). Within Ray-Order, "Standard" refers to the method that does not employ subpixel repurposing, instead calculating the ray parameters required for the standard paradigm and subsequently rendering the necessary three times the pixel amount in a ray-order manner.}
\label{tab:nerf_comp}
\resizebox{\textwidth}{!}{%
\begin{tabular}{lllllllllllllllllllllllll}
\hline
 & \multicolumn{12}{c|}{Synthetic Blend} & \multicolumn{12}{c}{Mip-NeRF360} \\ \hline
 & \multicolumn{6}{c}{Instant-NGP} & \multicolumn{6}{c|}{SNeRG} & \multicolumn{6}{c}{Instant-NGP} & \multicolumn{6}{c}{SNeRG} \\ \hline
 & \multicolumn{2}{c}{7.9-inch 2K} & \multicolumn{2}{c}{15.6-inch 4K} & \multicolumn{2}{c}{65-inch 8K} & \multicolumn{2}{c}{7.9-inch 2K} & \multicolumn{2}{c}{15.6-inch 4K} & \multicolumn{2}{c|}{65-inch 8K} & \multicolumn{2}{c}{7.9-inch 2K} & \multicolumn{2}{c}{15.6-inch 4K} & \multicolumn{2}{c}{65-inch 8K} & \multicolumn{2}{c}{7.9-inch 2K} & \multicolumn{2}{c}{15.6-inch 4K} & \multicolumn{2}{c}{65-inch 8K} \\
 & \multicolumn{1}{c}{RMSE} & \multicolumn{1}{c}{FPS} & \multicolumn{1}{c}{RMSE} & \multicolumn{1}{c}{FPS} & \multicolumn{1}{c}{RMSE} & \multicolumn{1}{c}{FPS} & \multicolumn{1}{c}{RMSE} & \multicolumn{1}{c}{FPS} & \multicolumn{1}{c}{RMSE} & \multicolumn{1}{c}{FPS} & \multicolumn{1}{c}{RMSE} & \multicolumn{1}{c|}{FPS} & \multicolumn{1}{c}{RMSE} & \multicolumn{1}{c}{FPS} & \multicolumn{1}{c}{RMSE} & \multicolumn{1}{c}{FPS} & \multicolumn{1}{c}{RMSE} & \multicolumn{1}{c}{FPS} & \multicolumn{1}{c}{RMSE} & \multicolumn{1}{c}{FPS} & \multicolumn{1}{c}{RMSE} & \multicolumn{1}{c}{FPS} & \multicolumn{1}{c}{RMSE} & \multicolumn{1}{c}{FPS} \\ \hline
\multicolumn{25}{l}{Multi-View} \\ \hline
LR & 8.283  & 4.425 & 8.166  & 2.532 & 7.575 & 0.509 & 9.821  &11.721  & 9.410  &4.065  & - &  \multicolumn{1}{l|}{1.823} &10.596 & 0.657 & 10.240 & 0.331 & 9.770 & 0.052 & 12.498 &7.067  & 11.894 &2.462 & - &0.447  \\
MR & 5.285  & 1.965  &4.718  & 0.574 & 4.006 & 0.094 & 6.094 &9.601 & 5.552 &2.519  & - &  \multicolumn{1}{l|}{-} & 5.869 & 0.210  & 5.037 & 0.083 & 4.179 & 0.013  & 7.845 &1.988  & 7.144 & 0.654  & - & -  \\
HR & 0.000 & 0.576  &0.000  & 0.150 & 0.000 & 0.023 & 0.000 &4.238  & 0.000 &0.859 & - &  \multicolumn{1}{l|}{-} & 0.000 & 0.053 & 0.000 & 0.021 & 0.000 & 0.003 & 0.000 &0.521  & 0.000 & 0.155 & - & -  \\ \hline
\multicolumn{25}{l}{Ray-Order} \\ \hline
Standard & 0.000 &\colorbox{pink!30}{9.709}  & 0.000 & \colorbox{pink!30}{2.967} & 0.000 & \colorbox{pink!30}{0.728} & 0.000 &\colorbox{pink!30}{12.305}  & 0.000 &\colorbox{pink!30}{4.163} & - &  \multicolumn{1}{l|}{-} & 0.000 & \colorbox{pink!30}{0.867} & 0.000 & \colorbox{pink!30}{0.418} & 0.000 & \colorbox{pink!30}{0.093} & 0.000 &\colorbox{pink!30}{8.290}  & 0.000 &\colorbox{pink!30}{2.954}  & - & - \\
Ours & - & \colorbox{purple!30}{18.519}  & - & \colorbox{purple!30}{6.803} & - & \colorbox{purple!30}{1.553} & - &\colorbox{purple!30}{25.406}  & - &\colorbox{purple!30}{8.402}  & - &  \multicolumn{1}{l|}{-} & - & \colorbox{purple!30}{1.802} & - & \colorbox{purple!30}{0.942} & - & \colorbox{purple!30}{0.227} & - &\colorbox{purple!30}{16.927}  & - &\colorbox{purple!30}{6.198}  & - & -  \\ \hline
\end{tabular}%
}
\end{table*}
\begin{table*}[!t]
\caption{Quantitative comparison of our method with the standard paradigm leveraging 3DGS. Within Multi-View, "Required Only" denotes the method that calculates pixel colors solely for the necessary locations during the final alpha-blending step of rasterization.}
\label{tab:gaussian_comp}
\resizebox{0.8\textwidth}{!}{%
\begin{tabular}{lllllllllllll}
\hline
 & \multicolumn{6}{c|}{Synthetic Blend} & \multicolumn{6}{c}{Mip-NeRF360} \\ \hline
 & \multicolumn{2}{c}{7.9-inch 2K} & \multicolumn{2}{c}{15.6-inch 4K} & \multicolumn{2}{c|}{65-inch 8K} & \multicolumn{2}{c}{7.9-inch 2K} & \multicolumn{2}{c}{15.6-inch 4K} & \multicolumn{2}{c}{65-inch 8K} \\
 & RMSE & FPS & RMSE & FPS & RMSE & \multicolumn{1}{l|}{FPS} & RMSE & FPS & RMSE & FPS & RMSE & FPS \\ \hline
\multicolumn{13}{l}{Multi-View} \\ \hline
LR & 10.88 & 11.915 & 10.87 & 6.279 & 6.18 & \multicolumn{1}{c|}{2.394} & 27.23 & 5.659 & 28.14 & 3.129 & 19.42 & 1.387 \\
MR & 4.81 & 8.049 & 3.71 & 2.977 & 2.04 &  \multicolumn{1}{c|}{0.624} & 15.26 & 3.582 & 11.86 & 1.627 & 5.83 & 0.325 \\
Required Only & 0.00 & 4.222 & 0.00 & 1.292 & 0.00 &  \multicolumn{1}{c|}{0.153} & 0.00 & 2.692 & 0.00 & 0.695 & 0.00 & 0.097 \\
HR & 0.00 & 3.112 & 0.00 & 0.974 & 0.00 &  \multicolumn{1}{c|}{0.118} & 0.00 & 1.957 & 0.00 & 0.518 & 0.00 & 0.071 \\ \hline
\multicolumn{13}{l}{Ray-Order} \\ \hline
Standard & -  &2.119  & -  & 0.841 & -  & \multicolumn{1}{c|}{0.104} & -  & 1.288 & -  & 0.441 & -  &0.061  \\
Ours & - & 4.574 & - & 1.852 & - & \multicolumn{1}{c|}{0.289} & - & 2.685 & - & 0.973 & - & 0.139\\ \hline
\end{tabular}%
}
\end{table*}

\section{Experiments}
In this section, we provide comprehensive experimental results to validate the effectiveness of our proposed DirectL method compared to the standard paradigm and further evaluate our algorithm through ablation studies.  Additionally, we demonstrate the enrichment and convenience in 3D display content brought by applying DirectL to a broader range of downstream tasks.

\subsection{Experimental Setup and Implementation}
We conduct experiments on three light field displays with varying parameters, detailed in Table \ref{tab:lfds} and Appendix \ref{Light Field Displays Prototype}. We \textbf{do not} utilize any SDKs or tools provided by the manufacturer. Our comparative analysis assesses DirectL against the standard MVR paradigm across different radiance fields. For NeRFs, we leverage the widely-adopted Instant-NGP \cite{muller2022instant} and the efficiently renderable SNeRG \cite{hedman2021baking} post baking. For 3DGS, we leverage the original 3DGS \cite{kerbl20233d}. Within the standard paradigm, we leverage the official CUDA implementations of Instant-NGP and 3DGS, as well as the WebGL implementation of SNeRG. For GPU-accelerated interlacing, we implement custom CUDA kernels for both the standard interlacing and the index-based array reordering in DirectL, with texture look-up implemented in WebGL. To achieve efficient ray-order rendering for NeRFs, we modify the original implementation of Instant-NGP and SNeRG. The ray-order 3DGS is implemented with CUDA, leveraging a compressed 8-wide BVH \cite{ylitie2017efficient} for efficient ray traversal. Each ray is mapped to a single CUDA thread. We reconstruct radiance fields using the Mip-Nerf360 \cite{barron2022mip} dataset and the synthetic Blender dataset \cite{mildenhall2021nerf}, training with official codes at the maximum supported resolution. Following \cite{fink2023efficient}, we set varying single-viewpoint resolutions—Low (LR), Medium (MR), and High (HR)—tailored for each light field display, as shown in Tab. \ref{tab:lfds}. LR represents a compromise on visual quality for real-time applications as recommended by manufacturers, while HR denotes the native screen resolution, and MR an intermediate value. For image quality evaluation, we use standard PSNR and SSIM metrics for 2D single-viewpoint images. For encoded images, using the HR encoded images from the standard paradigm as the ground truth, we measure the distance with RMSE following \cite{fink2023efficient}. It is important to note that while RMSE can reflect visual quality to some extent for the same subpixel mapping relationship, it is meaningless if the mapping differs. As no effective method currently exists to quantitatively assess the perceptual quality of encoded images \cite{fink2023efficient, shen2023virtual}, we conduct the user study with 30 unbiased participants. Although subjective, assessing visual quality from the user's perspective is more in line with the essence of 3D display: if it looks right, it is right. Participants rate each 3D image on three criteria—\textit{3D effect}, \textit{Clarity}, and \textit{Comfort}—each on a scale from 0 to 10 (more details in Appendix \ref{User Study}). The \textit{user-perception quality} score is empirically derived from three metrics with weights of (0.4, 0.3, 0.3). For each scene, we establish a sequence of camera poses and calculate the average time per frame over 200 frames. All experiments are performed on an Intel Xeon Silver 4110 CPU and a single NVIDIA RTX3090 GPU.

\begin{table*}[!t]
\caption{Results of user visual perception evaluation of DirectL versus standard paradigm leveraging SNeRG and Instant-NGP.}
\label{tab:user_snerg}
\resizebox{\textwidth}{!}{%
\begin{tabular}{ccccccccccccc}
\hline
 & \multicolumn{4}{c|}{7.9-inch 2K (SNeRG)} & \multicolumn{4}{c|}{15.6-inch 4K (SNeRG)} & \multicolumn{4}{c}{65-inch 8K (Instant-NGP)} \\
 & 3D Effect & Clarity & Comfort & \multicolumn{1}{l|}{Weighted Score} & 3D Effect & Clarity & Comfort & \multicolumn{1}{l|}{Weighted Score} & 3D Effect & Clarity & Comfort & Weighted Score \\ \hline
\multicolumn{13}{l}{Multi-View} \\ \hline
LR & 7.93 (238) & 7.40 (222) & 7.53 (226) & \multicolumn{1}{c|}{7.65} & 7.40 (222) & 6.93 (208)  & 7.03 (211) & \multicolumn{1}{c|}{7.15} & 7.47 (224) & 7.10 (213) & 7.07 (212) &  7.24\\
MR & \colorbox{pink!30}{8.00 (240)} & \colorbox{pink!30}{7.67 (230)} & \colorbox{purple!30}{7.63 (229)} & \multicolumn{1}{c|}{\colorbox{pink!30}{7.79}} & 7.60 (228) & 7.17 (215) & 7.27 (218) & \multicolumn{1}{c|}{7.37} & \colorbox{pink!30}{7.93 (238)} & 7.53 (226) & 7.60 (228) &  7.71\\
HR & \colorbox{purple!30}{8.07 (242)} & \colorbox{purple!30}{7.77 (233)} & \colorbox{purple!30}{7.63 (229)} & \multicolumn{1}{c|}{\colorbox{purple!30}{7.85}} & \colorbox{pink!30}{7.63 (229)} & \colorbox{purple!30}{7.43 (223)} & \colorbox{pink!30}{7.40 (222)} & \multicolumn{1}{c|}{\colorbox{pink!30}{7.50}} & \colorbox{purple!30}{8.13 (244)} & \colorbox{purple!30}{7.83 (235)} & \colorbox{pink!30}{7.87 (236)} &  \colorbox{purple!30}{7.96}\\ \hline
\multicolumn{13}{l}{Ray-Order} \\ \hline
Ours & \colorbox{pink!30}{8.00 (240)} & \colorbox{pink!30}{7.67 (230)}  & \colorbox{pink!30}{7.57 (227)} & \multicolumn{1}{c|}{7.77} & \colorbox{purple!30}{7.77 (233)} & \colorbox{pink!30}{7.37 (221)}  & \colorbox{purple!30}{7.47 (224)} & \multicolumn{1}{c|}{\colorbox{purple!30}{7.56}} & \colorbox{purple!30}{8.13 (244)} & \colorbox{pink!30}{7.73 (232)} & \colorbox{purple!30}{7.90 (237)} & \colorbox{pink!30}{7.94}\\ \hline
\end{tabular}%
}
\end{table*}
\begin{table*}[!t]
\caption{Results of user visual perception evaluation of DirectL versus standard paradigm leveraging 3DGS. In Multi-View, "HR w SR" refers to the method that interlaces using the mapping relationship obtained after subpixel repurposing following multi-viewpoint rendering.}
\label{tab:user_gau}
\resizebox{\textwidth}{!}{%
\begin{tabular}{ccccccccccccc}
\hline
 & \multicolumn{4}{c|}{7.9-inch 2K} & \multicolumn{4}{c|}{15.6-inch 4K} & \multicolumn{4}{c}{65-inch 8K} \\ 
 & 3D Effect & Clarity & Comfort & \multicolumn{1}{l|}{Weighted Score} & 3D Effect & Clarity & Comfort & \multicolumn{1}{l|}{Weighted Score} & 3D Effect & Clarity & Comfort & Weighted Score \\ \hline
\multicolumn{13}{l}{Multi-View} \\ \hline
LR & \colorbox{purple!30}{8.47 (254)} & 8.03 (241) & \colorbox{purple!30}{8.07 (242)} & \multicolumn{1}{c|}{\colorbox{purple!30}{8.22}} & 7.83 (235) & 7.20 (216) & 7.43 (223) & \multicolumn{1}{c|}{7.52} & 7.90 (237) & 7.33 (220) & 7.47 (224) &  7.60\\ 
MR & \colorbox{pink!30}{8.43 (253)} & 8.13 (244) & 7.87 (236) & \multicolumn{1}{c|}{\colorbox{pink!30}{8.17}} & 8.13 (244) & 7.77 (233) & \colorbox{purple!30}{7.67 (230)} & \multicolumn{1}{c|}{7.89} & 8.27 (248) & 7.87 (236) & 7.83 (235) & 8.02 \\
HR & 8.23 (247) & \colorbox{pink!30}{8.20 (246)} & \colorbox{pink!30}{7.97 (239)} & \multicolumn{1}{c|}{8.14} & \colorbox{pink!30}{8.37 (251)} & 7.77 (233) & \colorbox{pink!30}{7.57 (227)} & \multicolumn{1}{c|}{7.95} & \colorbox{pink!30}{8.50 (255)} & 8.07 (242) & \colorbox{purple!30}{7.97 (239)} & 8.21 \\
HR w SR & 8.33 (250) & \colorbox{pink!30}{8.20 (246)} & 7.90 (237) & \multicolumn{1}{c|}{8.16} & \colorbox{purple!30}{8.43 (253)} & \colorbox{pink!30}{7.90 (237)} & 7.50 (225) & \multicolumn{1}{c|}{\colorbox{purple!30}{7.99}} & \colorbox{purple!30}{8.63 (259)} & \colorbox{pink!30}{8.17 (245)} & \colorbox{pink!30}{7.93 (238)} & \colorbox{purple!30}{8.28} \\ \hline
\multicolumn{13}{l}{Ray-Order} \\ \hline 
Ours & 8.23 (247) & \colorbox{purple!30}{8.23 (247)} & 7.93 (238) & \multicolumn{1}{c|}{8.14} & 8.10 (243) & \colorbox{purple!30}{8.17 (245)} & \colorbox{pink!30}{7.57 (227)} & \multicolumn{1}{c|}{\colorbox{pink!30}{7.96}} & 8.43 (253) & \colorbox{purple!30}{8.30 (249)} & \colorbox{pink!30}{7.93 (238)} & \colorbox{pink!30}{8.24} \\ \hline
\end{tabular}%
}
\end{table*}

\begin{figure}[!t]
    \centering 
\includegraphics[width=0.7\columnwidth]{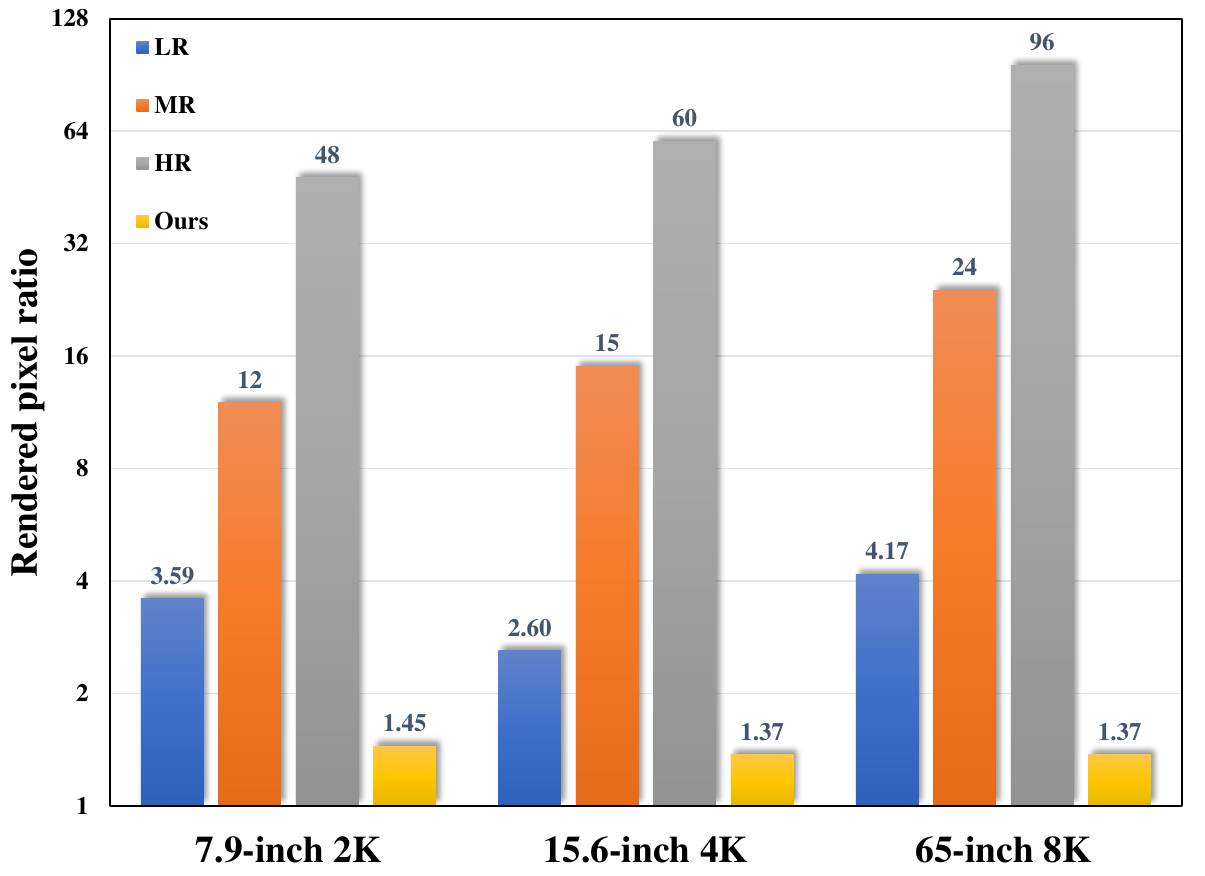}
    \caption{Comparison of the ratio of pixels rendered by different methods to the native screen resolution. For per-pixel rendering radiance field methods such as NeRFs, the number of rendered pixels directly impacts efficiency.}
    \label{fig:ratio}
\end{figure}

\subsection{Baseline Comparisons}
\label{Baseline Comparisons}
For fair comparison, we ensure that both NeRFs and 3DGS maximize the utilization of GPU cores within the memory limits during rendering. SNeRG faces memory overflow when rendering the MR and HR for the 65-inch 8K using WebGL. Although it is possible to render view-by-view, save, and then interlace, this process is extremely time-consuming (usually taking more than 10 minutes, which is no different from offline processing), hence we do not report the FPS and RMSE for this part. Due to the memory limitations of a single web page, the index used for reordering in 65-inch 8K cannot be loaded all at once, so we do not report this data either. Rendering an 8K resolution 2D image efficiently is already challenging; rendering a light field image for 65-inch 8K poses an even greater challenge, necessitating continuous iterative optimization in the future.  

Upon examining Fig. \ref{fig:ratio} and Tab. \ref{tab:nerf_comp}, the results demonstrate that DirectL significantly enhances the rendering efficiency of NeRFs on light field displays compared to the standard paradigm. Given the intrinsic per-pixel rendering support of NeRFs, the rendering time is strongly positively correlated with the number of pixels to be rendered. Through our meticulous analysis of the interlacing process and the pre-calculated ray parameters, even without utilizing the proposed subpixel repurposing, the \textit{Standard} method still outperforms all Multi-View methods in terms of FPS. It should be noted that for NeRFs, the light field images rendered by \textit{Standard} are identical to those of HR. Referring to the user perceptual quality scores for NeRFs in Tab. \ref{tab:user_snerg}, after subpixel repurposing, our method's Clarity and 3D Effect scores fall between MR and HR. The overall rating is slightly lower than MR on the 7.9-inch 2K display but is close to or better than HR on the other two screens. However, our method's rendering efficiency is an order of magnitude faster than HR on average. Observing the Clarity metric for LR, MR, and HR in Tab. \ref{tab:user_snerg}, it is found that the metric's level is highly consistent with the resolution of the single viewpoint, which is related to the extensive upsampling operations during interlacing. The NMSE metric in Tab. \ref{tab:nerf_comp} also reflects this issue to some extent.   

Upon reviewing the results in Tab. \ref{tab:user_gau} and Tab. \ref{tab:gaussian_comp}, the Clarity score of our implemented ray-order 3DGS achieves the highest in each group, with noticeable gaps between LR and MR compared to our method, particularly on higher-resolution screens. In overall scores, ray-order 3DGS consistently outperforms or equals HR, benefiting from ray-casting which produces sharper edges and enhances the 3D perception. However, due to the lack of tile-based parallel acceleration techniques, the rendering speed of ray-order 3DGS does not improve as significantly as NeRFs. Nevertheless, the substantial amount of redundant rendering in Multi-View somewhat compensates for this gap. Yet, ray-order 3DGS still offers rendering speeds at least 50\% faster than HR, and this is with maximizing GPU utilization by parallel rendering multiple viewpoints. If rendered serially view-by-view, HR would be slower.  

\begin{figure*}[!t]
    \centering
    \includegraphics[width=\textwidth, height=1.2\textwidth]{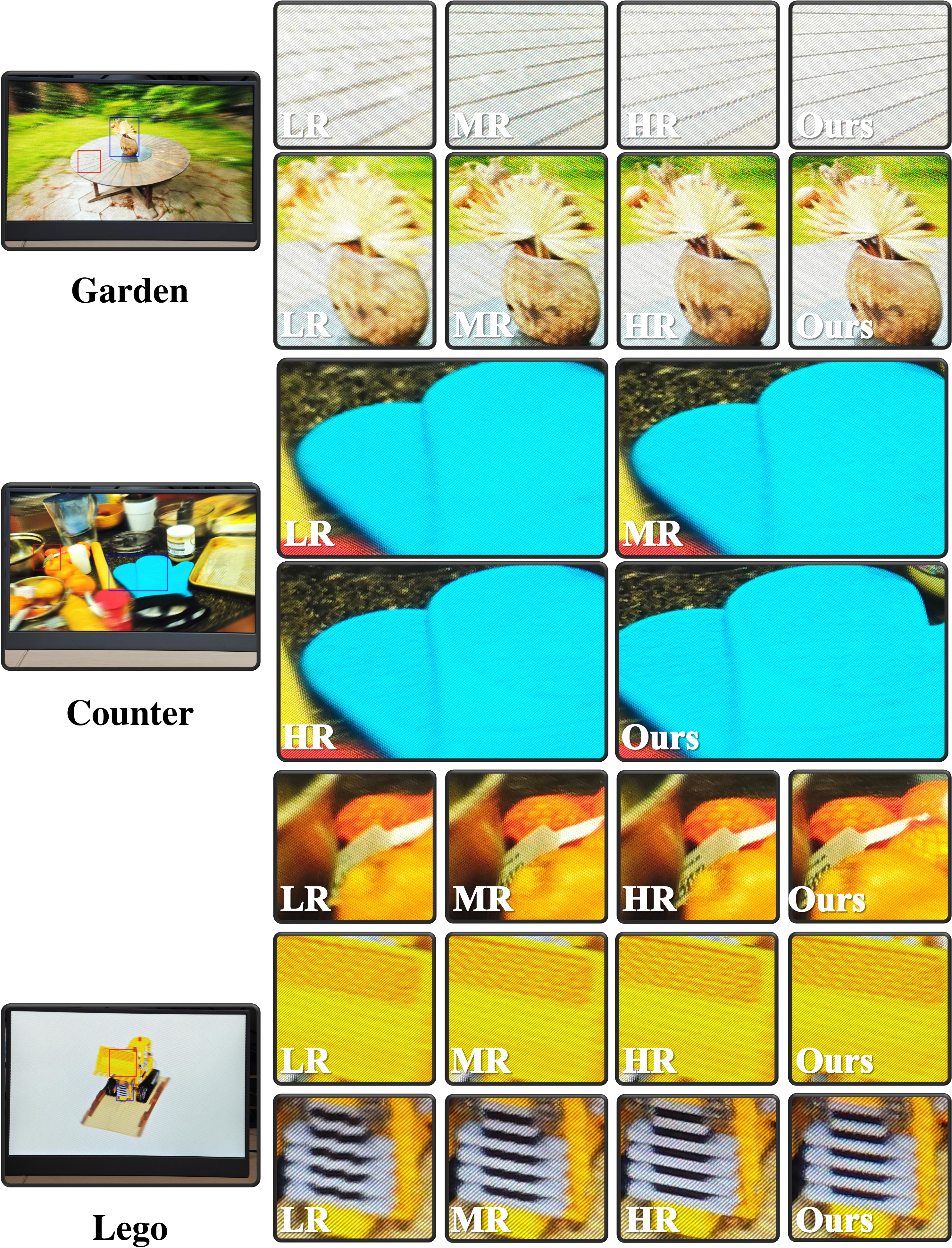} 
    \caption{Comparison of display effects between DirectL and standard paradigm. LR suffers from severe blurring due to extensive upsampling during interweaving, which also leads to the swelling of object edges. MR alleviates these issues to some extent but still encounters them. DirectL offers display quality nearly identical to HR, and in some cases, even superior. Due to real photography, perfect alignment of corresponding image positions is not always achievable.}     
    \label{fig:comp}
\end{figure*}

Fig. \ref{fig:comp} presents a visualization of 3d display effects of different methods under a single camera setup. Our method, along with HR, provides relatively the clearest display effects (the edges of the round table's wooden planks, the edges of the petals, the patterns on the gloves, the text on the tag, the texture of the Lego blocks). LR significantly degrades the display quality, while MR partially alleviates the issues. 

Upon analyzing the user study data from Tables \ref{tab:user_snerg} and \ref{tab:user_gau}, it is observed that 3DGS consistently scores higher than NeRFs across nearly all metrics, which aligns with the conclusion that 3DGS can reconstruct scenes more high-fidelity. Interestingly, on the 7.9-inch 2K display, the LR for 3DGS achieves the highest average weighted score. Following the user study, we interviewed several users who rated the LR highly. A consistent conclusion drawn from these interviews is that on the 7.9-inch screen, although the Clarity of LR is relatively lower, it appears more comfortable to watch and thus provides a stronger sense of depth protrusion.

\begin{table}[!t]
\caption{Rendered pixel ratio $\beta$ and the FPS of rendering encoded images under various $P_{\rm w}$ values by using Instant-NGP across the synthetic Blender dataset.} 
\label{tab:abl_subpixel_repurpose}
\resizebox{0.7\columnwidth}{!}{%
\begin{tabular}{c|cc|cc|cc}
\hline
            & \multicolumn{2}{c|}{7.9-inch 2K}  & \multicolumn{2}{c|}{15.6-inch 4K} & \multicolumn{2}{c}{65-inch 8K}     \\ 
                       & $\beta$  & FPS    & $\beta$    & FPS           & $\beta$     & FPS     \\ \hline
Standard             & 3.0000   & 9.709       & 3.0000           & 2.967      & 3.0000       & 0.728    \\ \hline
$P_{\rm w} = 1 $       & 1.4764   & 17.241        & 1.3701          & 6.711        & 1.5613      & 1.397    \\
$P_{\rm w} = 2 $       & 1.4467   & 18.519        & 1.3694          & 6.803        & 1.3708      & 1.553    \\
$P_{\rm w} = 3 $       & 1.3044   & 22.727        & 1.2868          & 7.576       & 1.2440      & 1.832     \\
$P_{\rm w} = 4 $       & 1.0790   & 24.390        & 1.1109          & 8.696       & 1.2222      & 1.835    \\ \hline
\end{tabular}%
}
\end{table}

\begin{table}[!t]
\caption{The impact of varying heap capacities $c_{\rm h}$ on visual quality and frames per second (FPS) in ray-order 3DGS compared with the original rasterization, using the Mip-NeRF360 dataset for rendering single-view images at a resolution of 1600 $\times$ 1060.}
\label{tab:abl_3dgs}
\resizebox{0.4\columnwidth}{!}{%
\begin{tabular}{lccc}
\\\hline
 & PSNR & SSIM & FPS \\\hline
$c_{\rm h}$ = 16 & 13.48 & 0.4519 & 15.04 \\
$c_{\rm h}$ = 32 & 21.66 & 0.6996 & 14.50 \\
$c_{\rm h}$ = 64 & 25.45 & 0.7874 &  13.64\\
$c_{\rm h}$ = 128 & 25.50 & 0.7888 & 12.27 \\
$c_{\rm h}$ = 256 & 25.57 & 0.7889 &  10.15\\
Original & 27.24 & 0.8164 & 93.07 \\\hline
\end{tabular}%
}
\end{table}

\begin{figure*}[!t]
    \centering
    \includegraphics[width=\textwidth]{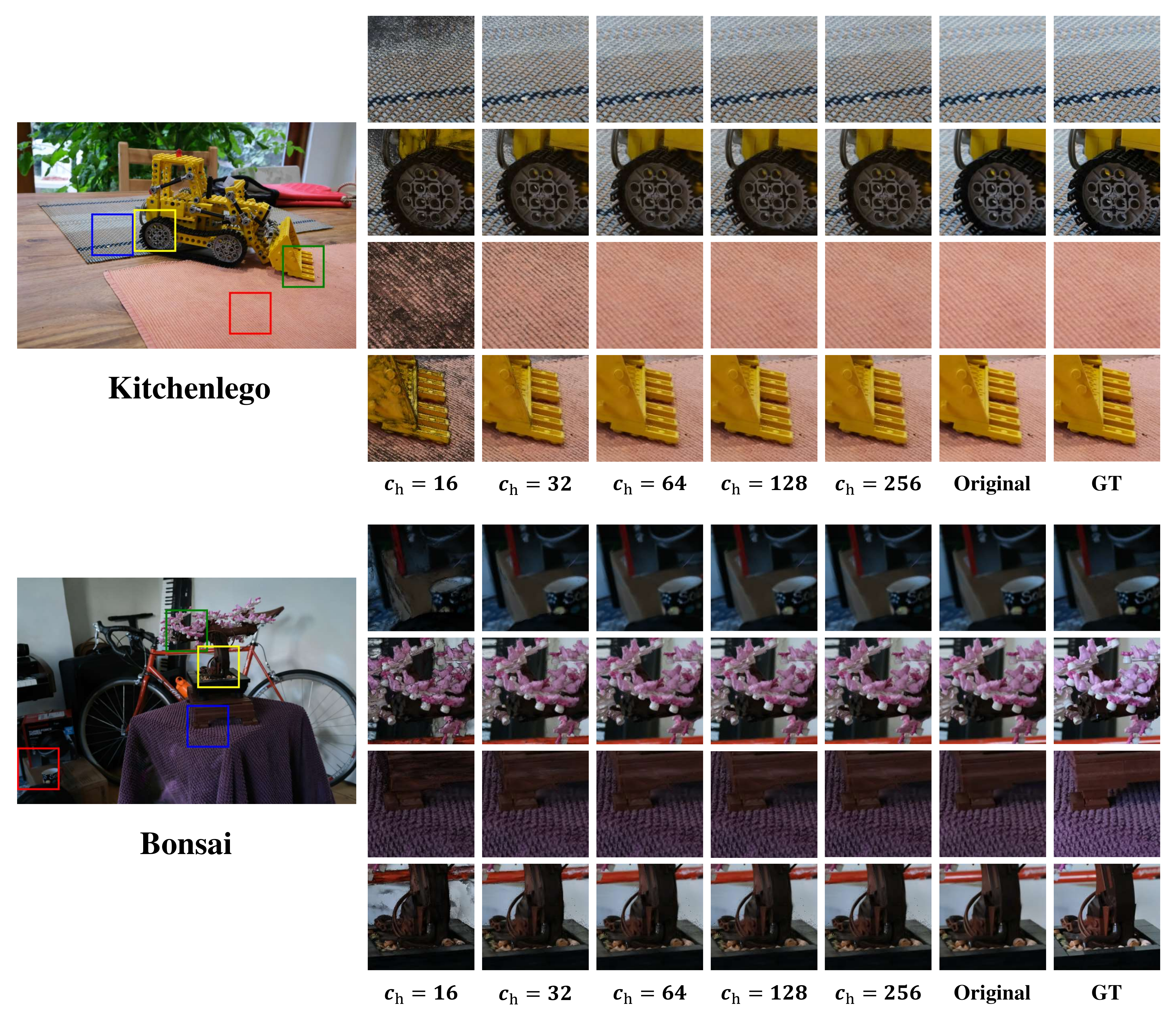}
    \caption{Visual comparison between our implemented ray-order 3DGS and the original rasterization. The parameter \( c_{\rm h} \) denotes the capacity of the max heap, which retains the \(c_{\rm h}\) Gaussians closest to the ray.}
    \label{fig:gaussian}
\end{figure*}
\begin{figure*}[!t]
    \centering
    \includegraphics[width=\textwidth]{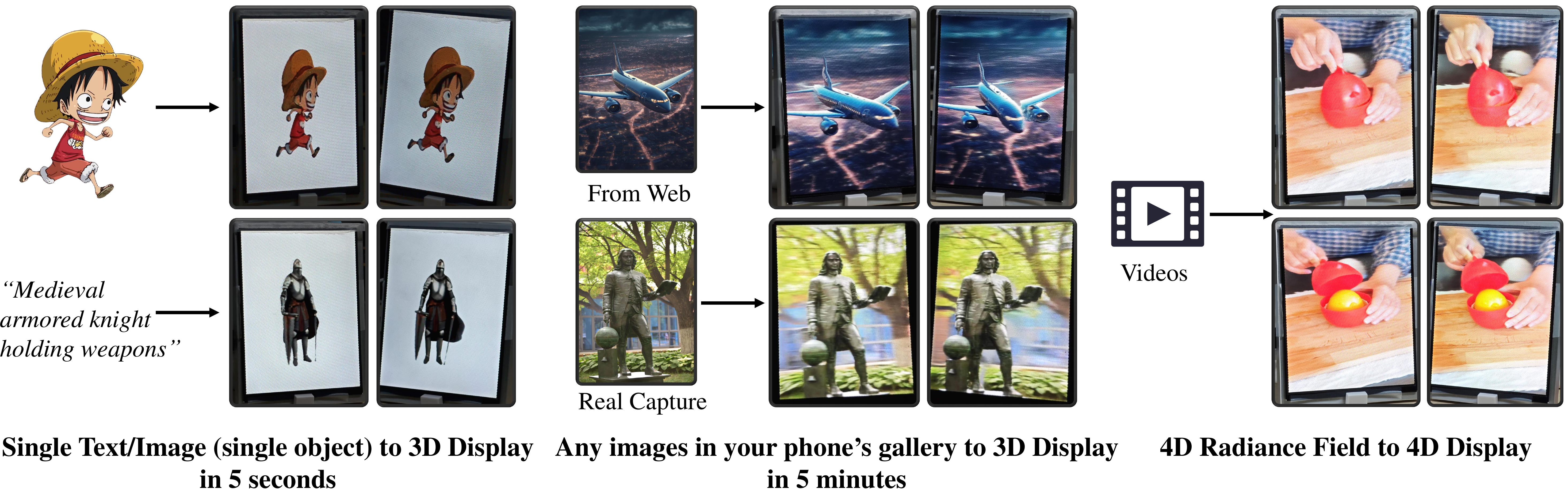}
    \caption{Integration of our proposed DirectL framework into downstream applications based on radiance fields. Demonstrating the mutual promotion and huge application prospects of the combination of the two fields.}
    \label{fig:app}
\end{figure*}
\subsection{Ablation Studies}
\label{abl}

In addition to the impact of certain modules analyzed in the previous section, this section specifically examines the influence of parameters within subpixel repurposing and ray-order 3DGS.

\textit{Search scope.}
In Alg. \ref{alg:resubpixl}, the LCD plane is divided into regions, each with $P_{\rm w}$ grating units. The size of each region, defined by $P_{\rm w}$, sets the search scope for repurposing unused subpixels from the same viewpoint, influencing the rendered pixel ratio $\beta$. This ratio, denoted as $\beta$, is the quotient of the pixels to be rendered over the native screen's pixel count. Tab. \ref{tab:abl_subpixel_repurpose} illustrates how an increased $P_{\rm w}$ per region corresponds to a reduced $\beta$. 
We explore the impact of $\beta$ on the rendering speed of light field images, as detailed in Tab. \ref{tab:abl_subpixel_repurpose}. Lower $\beta$ values increase rendering speed. However, a lower $\beta$ degrades display quality. We select $P_{\rm w} = 2$ to balance rendering speed with image quality. Our proposed subpixel repurposing method offers faster rendering of light field images compared to the standard correspondence mapping interlacing, without sacrificing visual quality.

\textit{Heap capacity.} The capacity of the Gaussian heap in ray-order 3DGS ($c_{\rm h}$) impacts rendering quality and speed. Each ray intersects fewer than 256 Gaussians on average, with potentially more in denser regions. As illustrated in Fig. \ref{fig:gaussian} and Tab. \ref{tab:abl_3dgs}, an low $c_{\rm h}$ leads to incomplete alpha-blending, resulting in significant quality degradation despite higher FPS. We set $c_{\rm h}=128$ as a trade-off when comparing with other methods. Since only one ray is sampled per pixel, discrete noise may arise, but we find that it does not impair the 3D display quality. On the contrary, rendering 3DGS in a ray-casting fashion enhances image sharpness, beneficial for clarity in 3D displays (see Tab. \ref{tab:user_gau}). While ray-order 3DGS single-view rendering is significantly slower than the original rasterization approach, akin to the performance gap between ray tracing and rasterization, this disparity is offset by the redundant computations in multi-view rendering, as discussed in the previous section.

\subsection{Applications}
In this section, we apply the DirectL framework to downstream tasks based on radiance fields. As demonstrated in Fig. \ref{fig:app}, the experimental results showcase an astonishing experience in 3D generation, reconstruction, and naked-eye 3D display, revealing the immense potential of combining radiance fields with light field displays.

\textbf{Single Text/Image to 3D Display: }
In our first application, we demonstrate how to manifest boundless imagination and cherished memories in the form of naked-eye 3D. Thanks to the learnable 3D representations of radiance fields and the flourishing development of generative models in recent years, we can easily convert imaginative text prompts into high-quality images using Stable Diffusion\cite{rombach2022high, podell2023sdxl, zhang2023adding, sauer2024fast}, and then transform single-view images into radiance field-based 3D representations using large reconstruction models\cite{xu2024grm, tochilkin2024triposr, hong2023lrm}, followed by the application of the DirectL framework for 3D display. For this application, we leverage Stable Diffusion v1-5 for text-to-image generation, TripoSR\cite{tochilkin2024triposr} for image-to-radiance field conversion, and finally, the DirectL framework for rendering. As shown in Figure \ref{fig:app}, users can now visualize a 360-degree, unobstructed, naked-eye 3D effect from a single-subject image or through text description within just 5 seconds.

People's phones contain photos that hold precious memories or are personally appealing downloaded from the internet. We aim to bring these memories to life in a 3D form. Building upon Neuralift\cite{xu2023neurallift}, we make improvements (focusing only on reconstruction and completion within the frontal $\theta_{\rm d}$ angle range, preserving the background) to display any 2D photo onto LFDs with high fidelity. As shown in Fig. \ref{fig:app}, users are capable of transforming any 2D image from their photo gallery, be it a family portrait or a landscape photograph, into a static light field image that fits within the viewable range of a 7.9-inch light field display.  

\textbf{4D Display:}
The fusion of 4D Radiance Fields \cite{wu20234d, xu20234k4d, yang2023deformable}with light field displays marks a leap in dynamic 3D content creation and visualization. This integration allows for the production of content that is complex in space and dynamic over time, closely mirroring real-world phenomena. After being encoded, the dynamic 3D content is display-ready for light field screens. By rendering the 4D radiance field onto the light field display, 3D display is equipped with the temporal dimension. Our used train data is sourced from HyperNeRF\cite{park2021hypernerf}. Currently, efficiently reconstructing high-fidelity 4D radiance fields remains a challenging task, often requiring specific video capture techniques. Translating arbitrary monocular videos from our gallery into high-fidelity 3D displays is extremely difficult due to computational costs and inter-frame consistency, yet it represents a meaningful and valuable area of research. We have also observed recent works\cite{gao2024cat3d, chu2024dreamscene4d, xu2024camco} demonstrating high 3D consistency, which aids in achieving this application. 

Another application with broad prospects in the cultural and tourism industry is the reconstruction and rendering of large-scale radiance fields \cite{liu2024citygaussian, kerbl2024hierarchical} on LFDs, enabling scene roaming.  
Similar applications already exist in VR, but leveraging LFDs allows for simultaneous viewing by multiple individuals without the need for wearable devices. We consider this application for future works.

\section{CONCLUSION}

In this paper, we first combine radiance fields and light field displays from display principles, proposing the DirectL paradigm to replace the existing standard paradigm. DirectL visualizes radiance fields in \textbf{Naked-eye 3D} efficiently. Extensive and comprehensive experiments demonstrate the superiority of DirectL for rendering NeRFs and 3DGS on light field displays. We also conduct a comprehensive visual quality assessment of the display effects from the user's perspective. Under the same display quality, DirectL is essentially an order of magnitude faster than the standard paradigm. We further leverage the proposed DirectL framework to showcase more downstream radiance field-based applications on light field displays. These applications greatly simplify the creation of 3D display content, enabling even non-professional users to easily create interesting and meaningful display content. Benefiting from the development of generative models and the DirectL paradigm, the entire creative process is also greatly shortened. All of these works undoubtedly demonstrate the tremendous application prospects of combining radiance fields with modern light field displays, paving a viable path for the commercialization of radiance field-related methods. We hope this study will attract more researchers to better combine these two fields and lead to more meaningful and imaginative applications!

\subsection{Limitations and Future works}

Beyond the work presented in our paper, the combination of these two fields remains largely unexplored, and we have taken only a modest initial step in this direction. We humbly and candidly acknowledge the following limitations and potential avenues for improvement to enhance the overall 3D display quality:

\begin{itemize}
    \item The FPS on 15.6-inch 4K and 65-inch 8K displays still falls significantly short of the real-time requirement. Due to the immense computational demands, real-time rendering of ultra-high-resolution light field images remains unattainable. Conventional 2D images can leverage super-resolution techniques, such as DLSS, to enhance display quality and FPS, but current super-resolution methods cannot handle light field encoded images (containing substantial high-frequency information). Super-resolving multiple 2D views would still result in inefficiency, making real-time performance impractical. A promising future direction is to investigate super-resolution for encoded images, specifically performing super-resolution in the ray space.
    \item The lack of quantitative metrics for assessing the visual perception quality of 3D displays. While numerous image quality evaluation metrics aligned with human visual perception exist for 2D images, there are no direct perceptual quality metrics for display effects in the 3D display domain. User studies provide an alternative approach but at a higher cost. 
    \item The pre-computed ray parameters and index matrices require more storage and memory space. Quantization and compression schemes could be explored for efficient storage and retrieval.
    \item The ray-casting efficiency of 3DGS is lower than rasterization. In Sec. \ref{3DGS}, we discuss some limitations and solutions for ray-order 3DGS. In addition to these, our future work will leverage hardware acceleration for ray tracing on modern GPUs \cite{parker2010optix, moenne20243d} to achieve more efficient ray-order 3DGS and consider implementing backpropagation in CUDA.
\end{itemize}

In summary, the convergence of these two fields necessitates sustained and in-depth exploration. The objectives of these two fields are congruent, both aiming to provide an enhanced display experience. Light field displays introduce a spatial dimension to conventional screens, while the radiation fields reconstruct scene depth from multiple 2D images. Our aspiration is that after a period of advancement, when engaging in video conversations with others, we can raise our phones and they will perceive live 3D imagery with the naked eye!

\bibliographystyle{ACM-Reference-Format}
\bibliography{sample-base}

\appendix
\section{Light Field Displays Prototype}
\label{Light Field Displays Prototype}
The light field display prototypes in this study are consistent with the display principle of \cite{ZXReal2024, LookingGlass2024a}, enhanced by sophisticated manufacturing techniques. However, the light field displays we used offer more flexible developability, which use the standard HDMI protocol to transmit display information. By employing unidirectional collimated light as the light source and leveraging multi-layered non-spherical grating composite fabrication to precisely control the optical path, our displays achieve an expansive viewing angle. The optimal surface shape of the cylindrical lens is derived through neural network iterative optimization, which takes into account both central and peripheral imaging quality, effectively reduces stray light, and controls optical distortion within wide-angle fields of view, ultimately achieving an impressive crosstalk reduction level of 1.25\%. This low crosstalk level, coupled with high definition, wide viewing angles, and substantial display depth, provides a remarkably lifelike visual experience, which necessitates rendering preferably at high resolution for a single viewpoint.  

\section{User Study}
\label{User Study}
We invite 30 unbiased participants to assess the visual perceptual quality of the 3D display, the majority of whom have not previously viewed a light field display. Each participant observes two sets of experiments: one rendered with NeRFs (7.9-inch 2K and 15.6-inch 4K are rendered by SNeRG, and 65-inch 8K is rendered by Instant-NGP, as SNeRG cannot render 65-inch's MR and HR online.) and the other with 3DGS. In each set, participants are randomly assigned a reconstructed scene to view the encoded images rendered by different methods. They are allowed to move freely within the observation area, with each image viewed for 30 seconds. After viewing each image, participants are asked to rate it on three criteria: \textit{3D effect}, \textit{Clarity}, and \textit{Comfort}, on a scale from 0 to 10, in integer steps. We explain the meaning of each criterion to the participants: \textit{3D effect} represents the perceived depth of the image, with 0 indicating no difference from a standard 2D screen, and 10 resembling the depth sensation of looking through a clean glass window at objects in a display case. \textit{Clarity} indicates the sharpness of the image, with 0 being indistinguishable and 10 comparable to a 1080p screen. \textit{Comfort} refers to the ease on the eyes during observation, with lower scores indicating discomfort (dizziness, soreness), and 10 being equivalent to viewing a 2D screen. Since most participants are new to light field displays, their initial impressions can be highly positive, potentially skewing their ratings for subsequent images. To mitigate this, we ensure that the first image each participant sees in each set is an HR image, with other methods presented randomly. Participants are not made aware of this arrangement. We encourage participants to evaluate the scores comprehensively from various viewing angles. After the evaluation, we randomly select 5 participants and, with their consent, film them in a surround video. We then use Instant-NGP to render their full bodies via DirectL on the light field display. Additionally, we allow them to choose a photo from their phone's gallery, lift it into 3D, and render it onto the light field display. After viewing the results, we interview them about their impressions of the entire process. The feedback is unanimously positive, and each participant expresses a keen interest in purchasing a product that can end-to-end display phone photos on the light field display.

\end{document}